\documentclass[acmsmall]{acmart}
\usepackage{algorithm}
\usepackage{color, colortbl}
\usepackage[noend]{algpseudocode}
\makeatletter
\usepackage{amsmath}
\usepackage{svg}
\usepackage{graphicx}
\def\algbackskip{\hskip-\ALG@thistlm}
\makeatother
\algnewcommand\algorithmicforeach{\textbf{for each:}}
\algnewcommand\ForEach{\item[ \algorithmicforeach]}
\algdef{S}[FOR]{ForEach}[1]{\algorithmicforeach\ #1\ \algorithmicdo}
\usepackage{caption}
\usepackage{subcaption}
\usepackage{array}
\usepackage{multirow}
\usepackage{array}
\newcolumntype{P}[1]{>{\centering\arraybackslash}p{#1}}
\usepackage{amsmath}

\newcommand\MyBox[2]{
  \fbox{\lower0.75cm
    \vbox to 1.7cm{\vfil
      \hbox to 1.7cm{\hfil\parbox{1.4cm}{#1\\#2}\hfil}
      \vfil}%
  }%
}
\AtBeginDocument{%
  \providecommand\BibTeX{{%
    \normalfont B\kern-0.5em{\scshape i\kern-0.25em b}\kern-0.8em\TeX}}}

\setcopyright{acmcopyright}
\copyrightyear{2018}
\acmYear{2018}
\acmDOI{10.1145/1122445.1122456}

\acmJournal{JACM}
\acmVolume{37}
\acmNumber{4}
\acmArticle{111}
\acmMonth{8}

\begin{document}

\title{Understanding COVID-19 Effects on Mobility: A Community-Engaged Approach}

 
\author{Arun Sharma}
\email{sharm485@umn.edu}
\orcid{0002-6908-6960}
\affiliation{
  \institution{University of Minnesota, Twin Cities}
  \city{Minneapolis}
  \state{Minnesota}
  \country{USA}
}

\author{Majid Farhadloo}
\email{farha043@umn.edu }
\orcid{0002-6908-6960}
\affiliation{
  \institution{University of Minnesota, Twin Cities}
  \city{Minneapolis}
  \state{Minnesota}
  \country{USA}
}

\author{Yan Li}
\email{lixx4266@umn.edu}
\orcid{0002-6908-6960}
\affiliation{
  \institution{University of Minnesota, Twin Cities}
  \city{Minneapolis}
  \state{Minnesota}
  \country{USA}
}

\author{Aditya Kulkarni}
\email{kulka262@umn.edu}
\affiliation{%
  \institution{University of Minnesota, Twin Cities}
  \city{Minneapolis}
  \state{Minnesota}
  \country{USA}
} 

\author{Jayant Gupta}
\email{gupta423@umn.edu}
\affiliation{%
  \institution{University of Minnesota, Twin Cities}
  \city{Minneapolis}
  \state{Minnesota}
  \country{USA}
} 

\author{Shashi Shekhar}
\email{shekhar@umn.edu}
\affiliation{%
  \institution{University of Minnesota, Twin Cities}
  \city{Minneapolis}
  \state{Minnesota}
  \country{USA}
}
\renewcommand{\shortauthors}{Arun Sharma, Majid Farhadloo, Yan Li, Aditya Kulkarni, Jayant Gupta and Shashi Shekhar}


\begin{abstract}
Given aggregated mobile device data, the goal is to understand the impact of COVID-19 policy interventions on mobility. This problem is vital due to important societal use cases, such as safely reopening the economy. Challenges include understanding and interpreting questions of interest to policymakers, cross-jurisdictional variability in choice and time of interventions, the large data volume, and unknown sampling bias. The related work has explored the COVID-19 impact on travel distance, time spent at home, and the number of visitors at different points of interest. However, many policymakers are interested in long-duration visits to high-risk business categories and understanding the spatial selection bias to interpret summary reports. We provide an Entity Relationship diagram, system architecture, and implementation to support queries on long-duration visits in addition to fine resolution device count maps to understand spatial bias. We closely collaborated with policymakers to derive the system requirements and evaluate the system components, the summary reports, and visualizations.


\end{abstract}



\maketitle

\section{Introduction}
The coronavirus disease 2019 (COVID-19) has impacted public health with hundreds of thousands of mortalities and millions of confirmed cases. COVID-19 policy interventions significantly changed people’s mobility patterns in many places, including increasing telecommuting and online shopping, as well as reducing urban trips and traffic congestion.

Given aggregated privacy-protected mobile device data, we aim to understand the impact of COVID-19 policy interventions on mobility. In collaboration with policymakers in Minnesota since Spring 2020 \cite{shekhar2020a, Yan2020, Sharma2020}, we have explored issues such as: Where are the potential hotspots of the hangouts (e.g., long duration visits)? How are these hotspots evolving? Is the public complying with policy interventions? Providing answers to these questions is important for not only for addressing the issue of safely reopening the economy but also devising new ways to reduce the spread of coronavirus disease. Hence, we design a community-engaged tool via close collaboration with end-users such as policymakers, healthcare, and transportation analysts.

The challenges of this problem are three-fold. First, we are required to work closely with end-users in order to understand their policy and socioeconomic related questions (e.g., closing and reopening of businesses, etc.) The second challenge concerns handling cross-jurisdictional variability given the choice and timing of interventions within a given geographic region. Policy intervention differs within sub-regions of a given geographic area. Hence, it is challenging to create a learning model that encompasses all spatial variability in a fine-geographic space (i.e., census tracts or block groups) thereby increasing geographic complexity. The third challenge is handling large data volume which involve several points of interest throughout the US. Other challenges are related to addressing data quality questions (e.g., unknown sampling bias, location privacy) posed by the end-users. 

Google \cite{GoogleData2020,GoogleCommunityMobility2020}, StreetLight \cite{streetLight2020} and other sources provide rich mobility location data for understanding changes in human mobility over time. They provide aggregated mobility data in the form of reports \cite{GoogleCommunityMobility2020} summarizing mobile device data. However, such reports are limited to the frequency of visits to certain points of interest (POIs) aggregated by certain geographic areas (e.g., cities, states, and countries) and do not consider visit duration. Hence, these reports do not separate short and long-duration visits, which are of interest to local and state governments towards reducing disease spread. Other platforms include early work based on an interactive dashboards for human mobility trends \cite{desjardins2020rapid, dong2020interactive, gao2020mapping, samet2020using}, which includes Geographic Information System (GIS) visualization at the county-level computing on the fly statistical measures. However, none of the application dashboards actively engage end-users to address complex queries and generate custom reports at finer geographic complexity. 

\vspace{0.2em}
\textbf{Contributions:}
Our main contributions are as follows:
\vspace{-\topsep}
\begin{itemize}
    \setlength{\parskip}{0pt}
    \item We describes a community-engaged COVID-19 decision support system, based on close collaboration with end-users, who provided the system requirements such as the queries of interest (e.g., business categories with many long-duration visits, sampling bias, data privacy safeguards, etc.)
    \item We present the Entity-Relationship diagram to improve understanding of the aggregated mobile device data and facilitates a richer set of queries such as those related to long duration visits.
    \item We report the user feedback on summary reports and visualizations generated by our system leveraging the Entity-Relationship diagram.
\end{itemize}
\vspace{-\topsep}

\textbf{Scope:} 
Our aim is to understand spatiotemporal patterns of mobility affected by COVID-19 in Minnesota using aggregated privacy-protected mobile device data. COVID cases correlation and detailed assessment of data quality beyond simple sanity checks and the characterizing relationship between disease spread and mobile-phone data fall outside the scope of this paper.

\textbf{Outline:} The rest of the paper is organized as follows: Section \ref{novelity} describes application domains and policy context. Section \ref{DecisionSystem} uncovers the overall system architecture with a brief description of each layer. Section \ref{ER_diagram} details the proposed entity-relationship diagram for weekly pattern data and identified in response to the questions from end-users. Section \ref{sec:Validation} describes the validation process used to verify the design of a new schema via a case study. Section \ref{user_feedback} discusses user feedback and data quality issues. Section \ref{related_work} gives a broad overview of related work. Lastly, Section \ref{C_FW} concludes this work and lists plans for future research.

\section{Application Domain and Policy Context} \label{novelity}
\begin{figure*}[!ht]
    \includegraphics[width=0.8\textwidth]{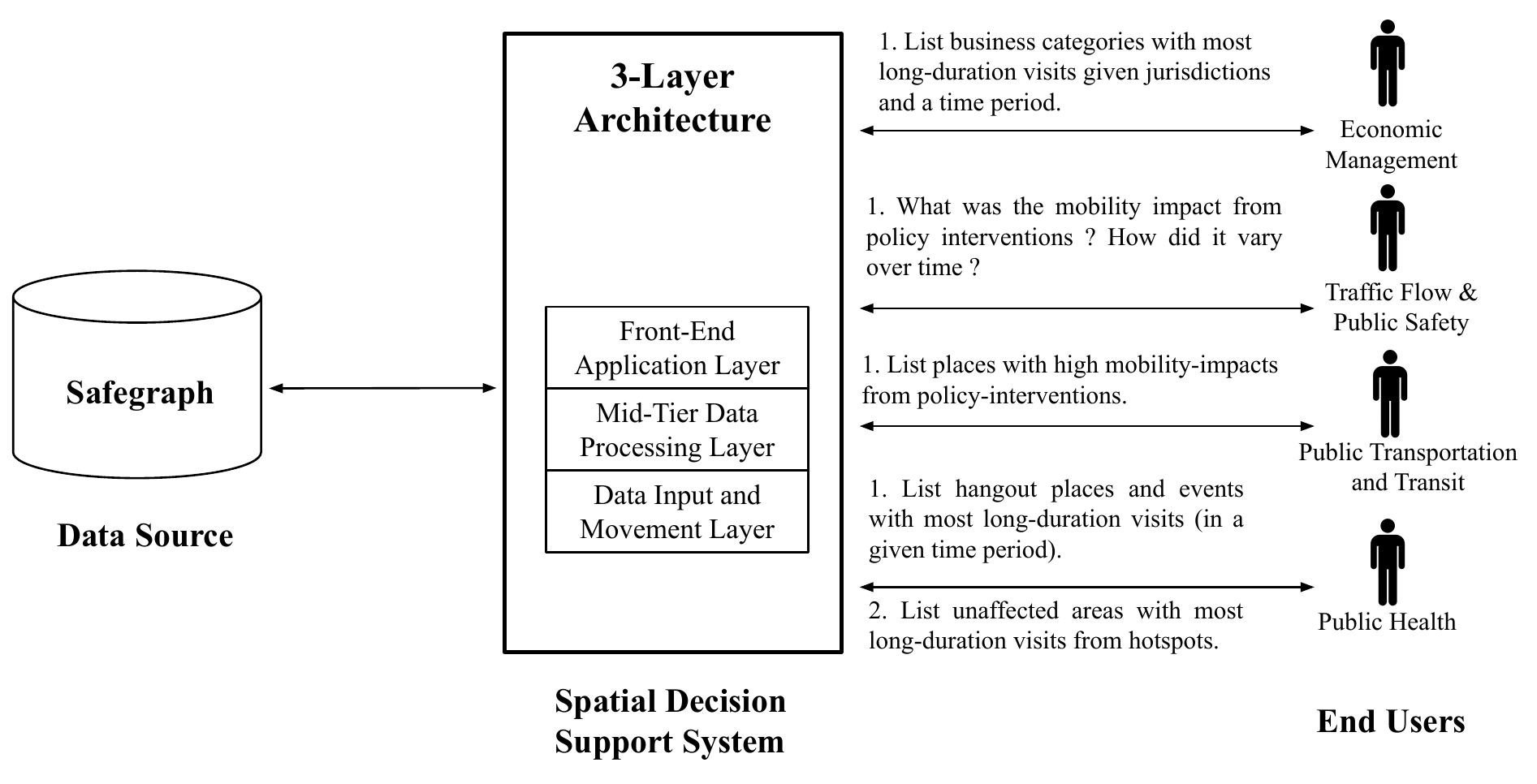}
    \caption{Context Diagram with end-users and data sources}
    \label{application_domain}
\end{figure*}

In March 2020, we were invited by policy analysts and the Center for Transportation Studies at the University of Minnesota and policy analysts to investigate a number of questions related to the impact of the pandemic and state-mandated interventions on mobility. In collaboration, we engaged with this audience through a sequence of online meetings, interviews, and email exchanges to better understand their information needs and understand the requirements for our decision-making system. The four main categories of end-users participating in these exchanges were researchers and policymakers in Public Health, Economic Management, Traffic Flow and Public Safety, and Public Transportation and Transit summarized in Figure \ref{application_domain}.

\textbf{Public Health:} Public health policymakers and researchers were interested in calibrating mobility-sensitive parameters in disease transmission dynamics models (e.g., SEIR\cite{bjornstad2020seirs} model). The mobility information of interest to this group included the size of gatherings and the number of long-duration visits at vulnerable places and events (e.g., indoor hangouts, high-density outdoor gatherings, such as super spreader events) since these may help estimate the number of contacts. In addition, they were interested in travel between disease hotspots and other areas to estimate the probability of future spread to new geographic locations.


\textbf{Economic Management:} Economic management policymakers and analysts were interested in understanding the impact of the policy interventions as well as how to safely reopen the economy. They asked for reports on the number of visits and the number of long-duration visits for different business categories (e.g., bars, full-service restaurants, and limited-service restaurants). They also requested an aggregate reports on trends in average distance traveled and average time spent at home in order to assess compliance such as Minnesota Stay-at-Home orders and phased reopening \cite{calendar2020} of the economy.

\textbf{Traffic Flow and Public Safety:} The traffic flow and public safety community were interested in understanding the current and future impact of the pandemic on travel demand (e.g., vehicle miles traveled for commuting and delivery of goods and services, etc.), quality of road service (e.g., congestion, average speed), safety (e.g., traffic accidents) and the environmental impacts (e.g., emissions). They also asked questions on the quality of data in terms of sampling bias, geographic coverage, sampling frequency, and comparison with the ground truth dataset (e.g., travel surveys and loop detectors). Furthermore, they also requested a deep engagement by inviting us to present our findings at technical conferences, public webinars, and legislative hearings for the State House of Representatives, Transportation, Finance, and Policy Committees.

\textbf{Public Transportation and Transit:} Analysts from transit were interested in most frequently visited Points of Interest (e.g., Minneapolis-St.Paul Airport) during the pandemic phase. This may help to plan new major bus routes surrounded by such POIs and their usage during the stay-at-home order or reopening phase. In addition, they were also interested in understanding mobility patterns based on finer temporal granularity (e.g., commute hours) for certain days in the week (e.g., Weekdays vs Weekends) and different calendar events.

\section{Community-Engaged Decision Support System}
\label{DecisionSystem}
This section provides a brief overview of the proposed decision support system developed in consultation with our end-users. The system generates weekly reports based on policy intervention questions posed by the end-users during the course of the pandemic. The end-users from interdisciplinary fields can interact with our decision support system and pose questions of interest (e.g., long duration visit queries). The system outputs a customized report along with supplemented details for answering data quality concerns (e.g., sampling bias, differential privacy). 

The system has three layers: a Data Storage and Movement Layer, a Mid-Tier Data Processing Layer, and a Front-end Application Layer. Details related to each layer are as follows:

\textbf{Data Storage and Movement Layer:} The first layer consists of three components: External Data Sources, Data Warehouse Servers, and the Client. The primary goal of this layer is gathering mobility data from remote geospatial servers to the client and mid tier Data Processing Layer. The Data Source Layer contains source files such as Safegraph that are periodically fetched by the Data Warehouse Servers and saved in a COVID-19 Data Warehouse. The COVID-19 Data Warehouse is timely maintained by USpatial \cite{Uspatial}, which periodically uploads comma-separated variables files to their PostgreSQL servers across different tables (e.g., weekly patterns and social distancing). These records are fetched by the client using secured VPN performing Extract Transform Load (ETL) operations and saved in a relational database table. The data is then prepared for more complex analysis based on policy intervention questions, as discussed in the Mid-Tier Data Processing Layer.

\begin{figure*}[!ht]
    \includegraphics[width=0.8\textwidth]{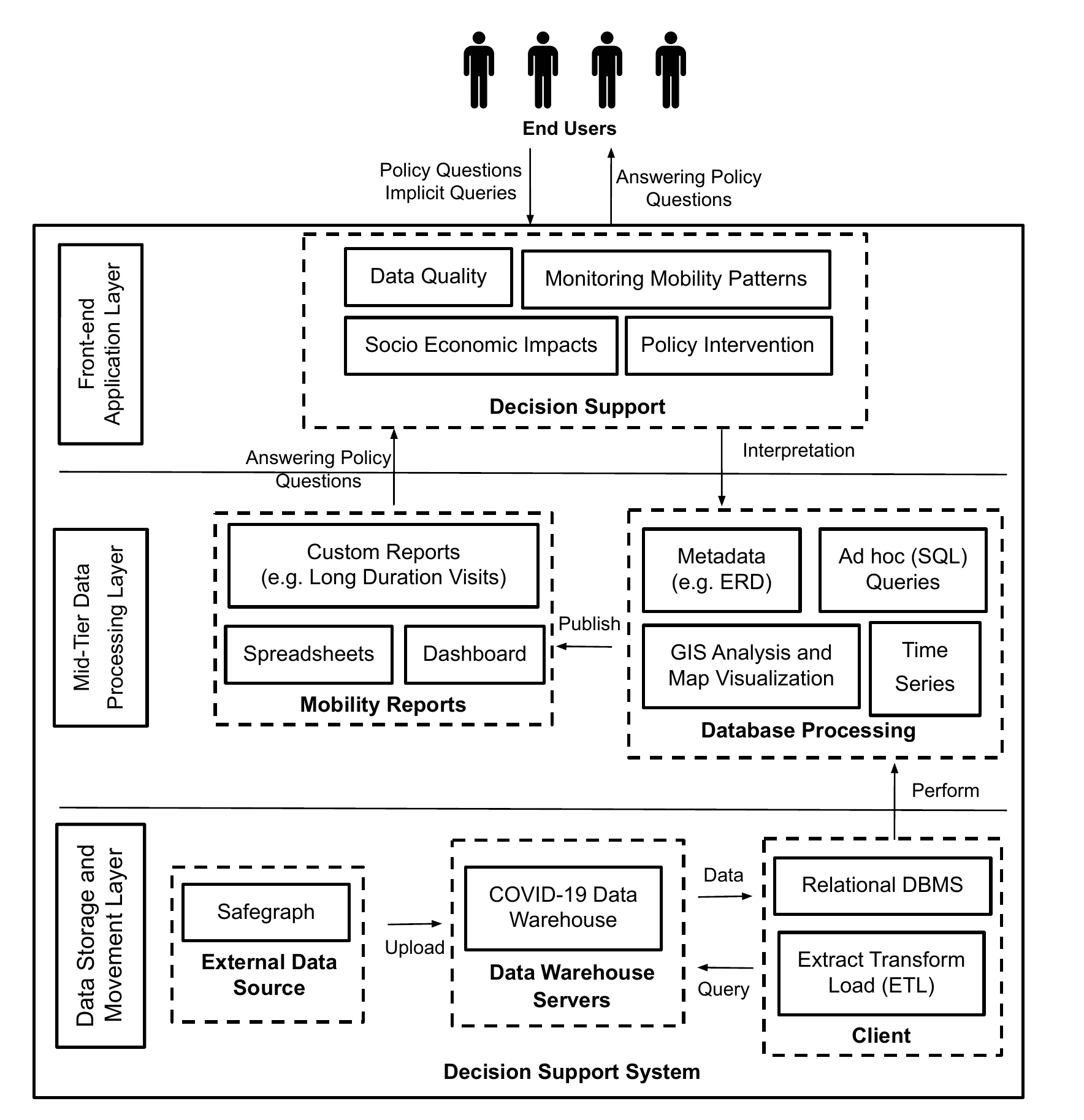}
    \caption{3-Layer Architecture of the Community-Engaged Decision Support System}
    \label{Architecture}
\end{figure*}

\begin{figure*}[!ht]
    \centering
    \includegraphics[width=0.5\textwidth]{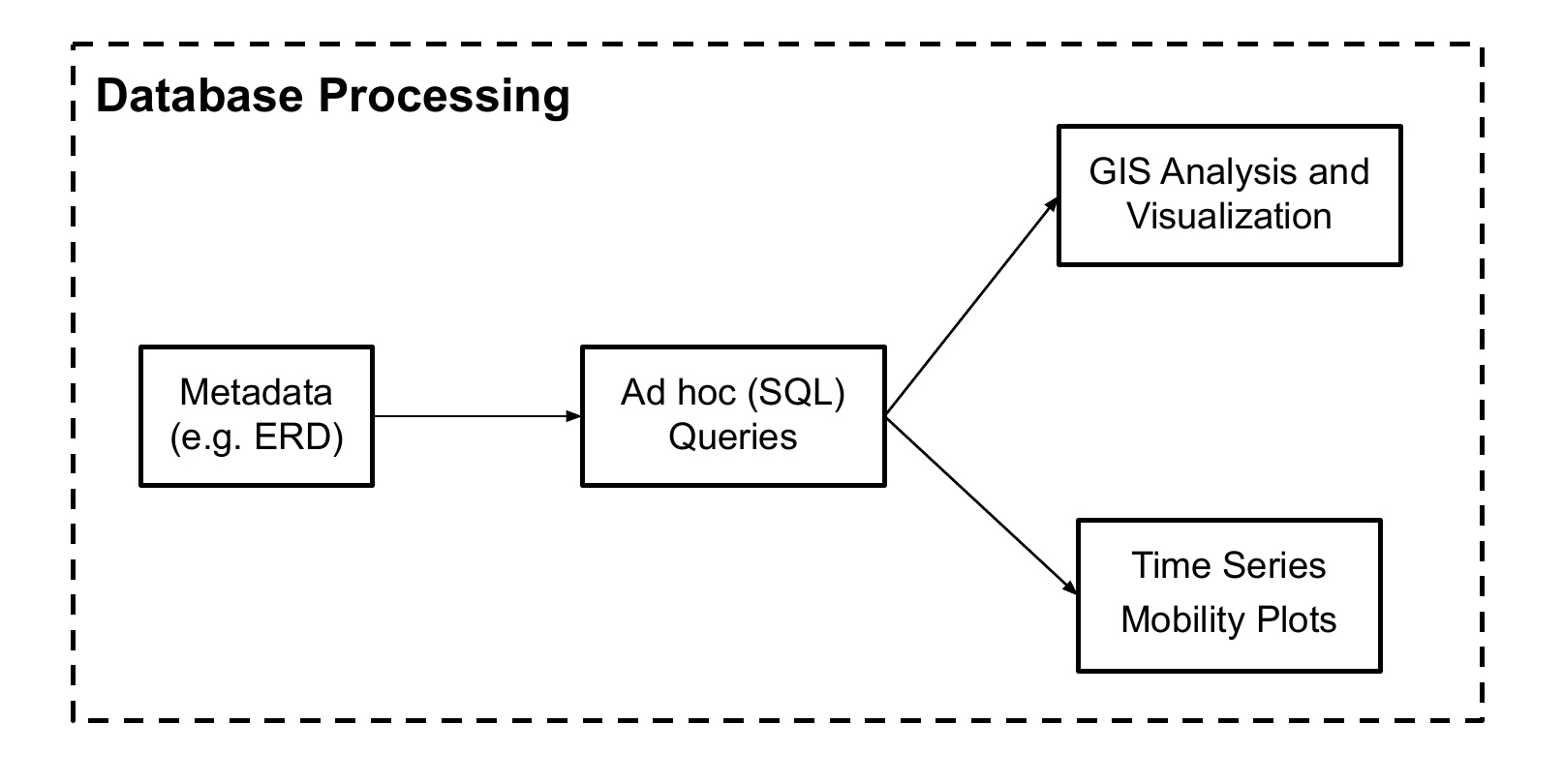}
    \caption{Database Processing Component \textcolor{blue}{(Best in Color)}.}
    \label{Sub-Architecture}
\end{figure*}

\textbf{Mid Tier Data Processing Layer:} The main objective of Mid-tier data processing is to perform desired spatial and spatiotemporal mobility analytics based on a given pre-processed dataset from the client. The layer has two components, Database Processing, and Mobility Reports. Within Database Processing component we first check if policy intervention questions can be answered via an Entity Relationship Diagram. If yes, then the questions can be further formalized via SQL queries. These queries may require additional I/O operations from the data warehouse servers in the Data Storage Layer. After formalizing and executing ad-hoc SQL queries, we perform more detailed mobility patterns via time-series plots and other geographic maps and visualizations (shown in Figure \ref{Sub-Architecture}). Based on the policy questions, such queries further involve join operations with different tables in Safegraph for providing Time Series Mobility Analysis. These findings are then documented in the form of spreadsheets and presentations to communicate with policymakers in the interpretation of mobility patterns.

\textbf{Front End Application Layer:} The primary objective of this layer is to provide decision support and interpretation of the results published in the form of reports the end-users. In addition, end-users can pose questions and queries (e.g., long-duration visits) and evaluate policy decisions based on human mobility patterns. These reports are further verified for accordance with the policy intervention calendar \cite{calendar2020} of Minnesota.

\section{Entity Relationship Diagram for Weekly Pattern Data on Safegraph Dataset}\label{ER_diagram}
\begin{figure*}[!ht]
    \centering
    \includegraphics[width=1.0\textwidth]{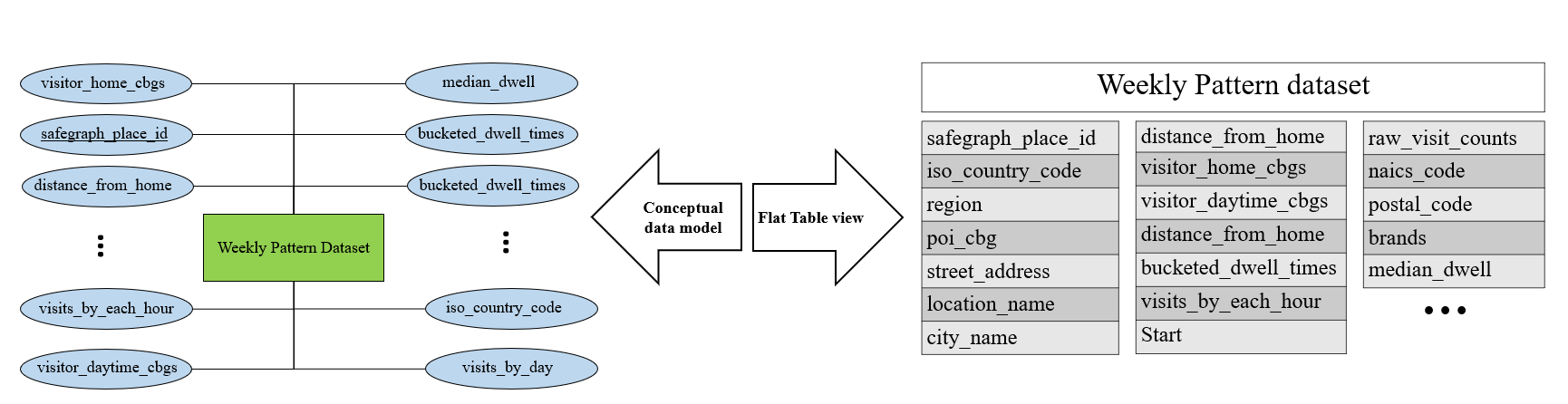}
    \caption{Denormalized Schema of Weekly Patterns Data \textcolor{blue}{(Best in Color)}}
    \label{Denormalized}
\end{figure*}

SafeGraph \cite{safeGraph2020} is a mobility data vendor company that provides anonymized aggregated location data from mobile devices. In 2020, the company made it's data free to researchers studying the effects of COVID-19 on people's travel patterns to millions of Points Of Interests (POIs) around us (more details in Section \ref{dataset}). Even though Safegraph \cite{safeGraph2020} provides invaluable data, its denormalized (a single flat-table) schema makes it hard to discern the semantic richness of the data and and support ad-hoc queries such as those related to long-duration visits. Figure \ref{Denormalized} shows the denormalized schema of the "weekly pattern" table from the SafeGraph dataset. With this format, it is hard to answer queries that distinguish between frequently visited places (e.g., grocery stores, fast-food restaurants) and hangout places with long-duration visits (e.g., bars and full-service restaurants), a critical requirement for this study. For example, even though the number of visits to limited-service restaurants (e.g., fast food restaurants) is much greater visits the number of visits to alcoholic bars, bars are found to be more important in the transmission of COVID-19 \cite{bars1}.

To overcome these limitations, we designed a conceptual schema (Entity-Relationship diagram) that satisfies the 3rd normal form rule to support a richer set of ad-hoc queries for the currently available weekly pattern data from the Safe-Graph dataset. We initially asked data engineers at SafeGraph whether a conceptual data model (e.g., entity relationship diagram) existed on the current design of the database. Despite relatively complicated and inflexible design, no effort had previously been made towards building a conceptual data model. We also carried out an extensive search of recently published works by organizations and academic researchers involved in the SafeGraph COVID-19 Data Consortium \cite{safeGraph2020} but could not find any conceptual data model. Defining data conceptually improves understanding of the semantics of the data and enhances the interpretability of the data to a more general audience. It also facilitates representation of a richer set of queries by defining relational tables in 3rd normal form. Therefore, we designed an Entity-Relationship (ER) diagram that satisfies 3rd normal forms on the weekly pattern data from the SafeGraph dataset (Figure \ref{ERD}).

\subsection{Proposed Schema on Weekly Patterns Dataset}
In the early stage of the Entity Relationship (ER) diagram design, we encountered many challenges in specifying the major entities. In current dataset, the mobile device data is currently aggregated at a Census Block Group level to address privacy concerns and data protection. Furthermore, a data redundancy issue arises since various data attributes are replicated across several key-value pairs that make it hard to distinguish major entities.

\begin{figure*}
    \centering
    \includegraphics[width=1.0\textwidth]{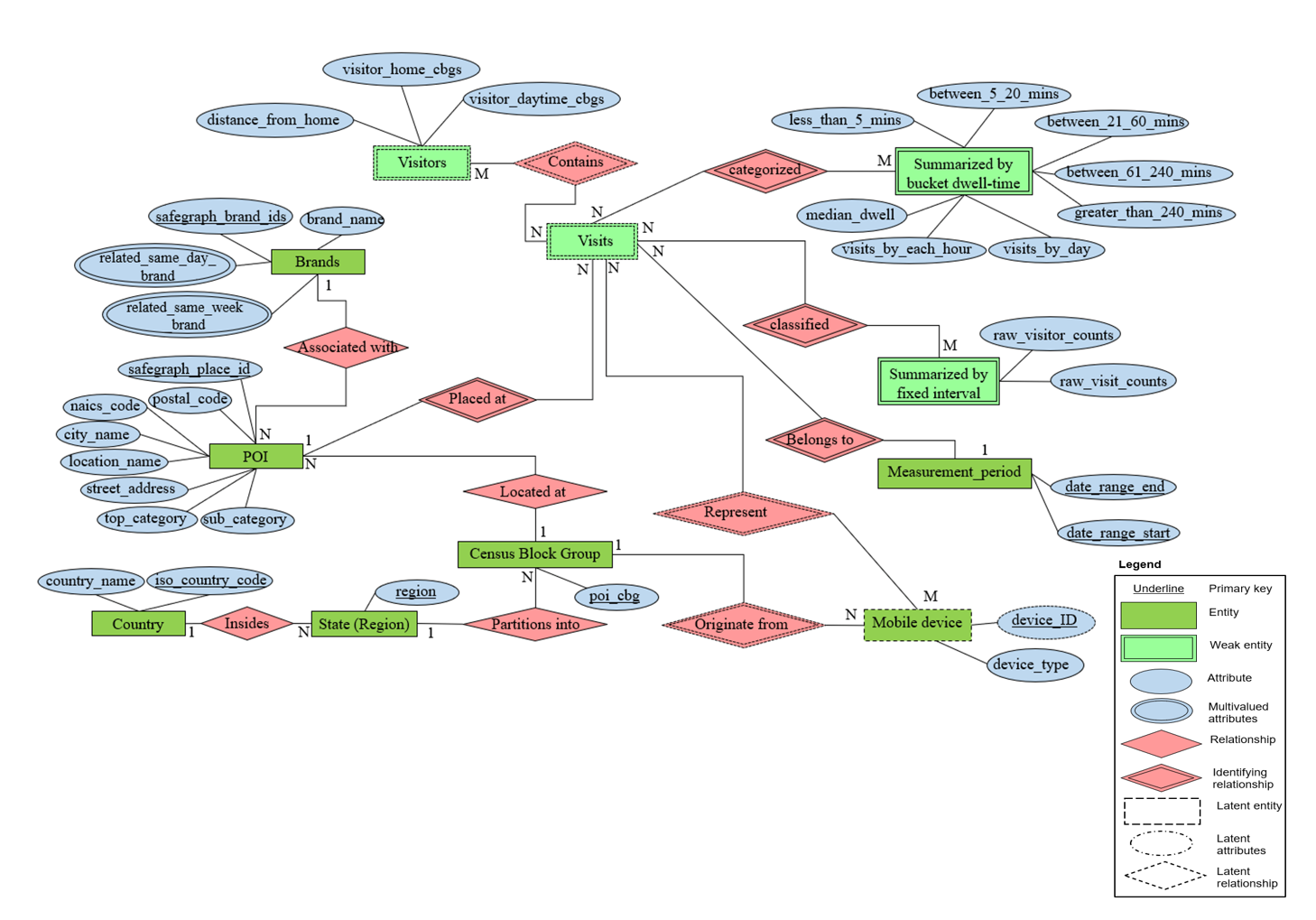}
    \caption{A Entity-Relationship diagram of the weekly pattern data from the SafeGraph dataset \textcolor{blue}{(Best in Color)}.}
    \label{ERD}
\end{figure*}

To address these issues, we have incorporated specific design decision steps that enable us to model ad-hoc queries (e.g., queries related to hangout places with many concurrent long-duration visits) in as much detail as possible to understand people's mobility affected by COVID-19. First, we consolidated the overall structure of the ER diagram by adding latent entities, such as ``Mobile device'', ``Visitors'', and ``Visits'' (green dashed green boxes in Fig. \ref{ERD}), which substantially improve our understanding of the denormalized schema. Due to privacy concerns as well as data protection, relevant information on latent entities are not currently available. However, having latent entities allows us to correctly identify entities from the available dataset and subsequently assign necessary existing or derived data attribute columns to each entity. We further improved the design of the ER diagram by summarizing the frequency of visits into bucket dwell times and fixed interval entities. For example, we model queries related to hangout places using bucket dwell time data attributes, including ``between\_21\_60\_mins'', ``between\_61\_240\_mins'', and ``greater\_than\_240\_mins''  from the ``Summarized by bucket dwell-time'' entity, which helps us address concerns from end-users regarding safely reopening of the economy. The complete list of entities used in designing the ER diagram is shown in Table \ref{entity-desc}.\\

\begin{table*}
\fontsize{6}{11}\selectfont
\centering
\caption{Entity name and description in the weekly pattern data from the SafeGraph dataset \textcolor{blue}{(Best in Color)}.}
\label{entity-desc}
\resizebox{\textwidth}{!}{%
\begin{tabular}{|l|l|}
\hline
Entity Name        & Description                                                           \\ \hline
Country            & The 2 letter ISO 3166-1 alpha-2 country code                         \\ \hline
State              & The state or territory                                                \\ \hline
Census Block Group & The census block group the POI is located within                      \\ \hline
Points of Interest (POI)        & The business categories that are located within the census block group                                              \\ \hline
Brands             & Brands that are associated with the POI                               \\ \hline
Measurement period        & The start and end time of the measurement period                      \\ \hline
Mobile device      & The mobile devices that are located within a census block group       \\ \hline
Visits             & The mobile devices that visit a POI at a certain time           \\ \hline
Visitors           & The summary statistics on visitors for certain POIs in the given time period\\ \hline
Summarized by bucket dwell-time & The summary statistics on visitors for certain POIs based on bucket dwell time (shorter, average, or longer) visits \\ \hline
Summarized by fixed interval    & The summary statistics on the visitors for certain POIs in a fixed interval                                         \\ \hline
\end{tabular}%
}
\end{table*}

\begin{table*}
\fontsize{9}{16}\selectfont
\centering
\caption{Relationship descriptions in weekly pattern dataset between pairs of entities \textcolor{blue}{(Best in Color)}.}
\label{entity-rels}
\resizebox{\textwidth}{!}{%
\begin{tabular}{|l|l|l|l|} 
\hline
Entity 1      & Relationship    & Entity 2                        & Description                                                                                                                \\ 
\hline
State         & Inside          & Country                         & A state or territory is located in a country                                                                               \\ 
\hline
State         & Partition into  & Census Block Group              & A state is spatially partitioned into multiple census block groups                                                         \\ 
\hline
POI           & Located at      & Census Block Group              & Multiple  POIs (i.e., business categories) are located in a census block group                                              \\ 
\hline
Brands        & Associated with & POI                             & One brand is associated with many POI                                                                                  \\ 
\hline
Mobile device & Originate from  & Census Block Group              & Multiple mobile devices originate from a census block group (i.e., visitors' home)  \\ 
\hline
Visits        & Placed at       & POI                             & Many visits are placed at a POI.                                                                                           \\ 
\hline
Visits        & Categorized     & Summarized by bucket dwell-time & Many visits are categorized by many bucket dwell-times (e.g., how long a visit lasts)                                       \\ 
\hline
Visits        & Classified     & Summarized by fixed interval    & Many visits are categorized by many fixed intervals.                                                                       \\ 
\hline
Visits        & Belongs to      & Measurement period                   & Visits to a POI from a mobile device belong to certain time and days of the week. \\ 
\hline Visits        & Contains    & Visitors                    & Related information (e.g., distance from home) from visitors are contained within visits.\\
\hline
Mobile device        & Represent    & Visits                    & Many  Multiple mobile devices represent visits are happened in a POI at given time. \\
\hline
\end{tabular}%
}
\end{table*}
Table \ref{entity-rels} relationships (reddish diamonds in Fig. \ref{ERD}) associate entities to one or more other entities. For example, a state is ``inside'' a country, each state ``partitions into'' many Census Block Groups (CBGs), and many Points Of Interests (POIs) are ``located at'' each CBG. Next, we incorporate a key design improvement into the ER diagram by adding latent identifying relationships (double dashed diamonds in Fig. \ref{ERD}) and derived identifying relationships (double diamonds in Fig. \ref{ERD}). Similar to identifying entities, having latent relationships in our ER diagram design enables us to go further in specifying the correct entities and data attribute columns for each entity. For example, mobile devices ``represent" visits that ``took placed at" a POI, which "belongs to" various times and days of the week. The number of visits can be further ``categorized/classified" as a fixed interval or bucket dwell time (e.g., visit duration longer than 20 minutes). Lastly, cardinality constraints (the numbers in Fig. \ref{ERD}) determine the least and greatest number of occurrences of an entity that could be related to a single occurrence of another entity. Cardinality constraints (e.g., one-to-many, many-to-many) represent business rules. For instance, we were able to create a hangout query place by modeling the frequency of visits that represent``many'' mobile devices, which ``originate from'' a CBG (e.g., home, workplace, etc.) and travel to another CBG. ``Many'' POIs (e.g., grocery store, full-service restaurants, bars) are also ``located'' at ``one'' CBG that is further ``associated'' with ``many'' brands (e.g., Walmart, McDonald's, Sally's). The complete list of the relationships between pairs of entities is shown in Table \ref{entity-rels}.

\subsection{An examination of the Entity Relationship Diagram}
A benefit of having a conceptual point of view Entity Relationship Diagram (ERD) is that it enables us to introduce a richer set of queries and subsequently assess which if any can be answered using the current dataset. Accordingly, we devised a set of 8 queries, 4 of which could be answered with the currently available dataset and 4 which require additional information or a supplementary dataset, as shown in Table \ref{query_questions}. 

\begin{table}\tiny
\centering
\caption{Query evaluation based on the proposed Entity Relationship Diagram \textcolor{blue}{(Best in Color)}}
\label{query_questions}
\begin{tabular}{|l|l|l|l|l|l|l|} 
\hline
End-user & Query & Evaluation & \begin{tabular}[c]{@{}l@{}}Implicit\\with \\current\\ schema\end{tabular} & \begin{tabular}[c]{@{}l@{}}Explicit\\ with \\current\\ schema\end{tabular} & \begin{tabular}[c]{@{}l@{}}Implicit\\with \\ERD\end{tabular} & \begin{tabular}[c]{@{}l@{}}Explicit\\ with \\ERD\end{tabular}  \\ 
\hline
\begin{tabular}[c]{@{}l@{}}Economic Management \\Public Transportation\\ and Transit\end{tabular} & \begin{tabular}[c]{@{}l@{}}1. What are the business categories located at a given\\Census Block Group and what is the distribution of \\visits in a given Census Block Group?\end{tabular} &\checkmark & \checkmark & x  & x & \checkmark \\  \hline
Economic Management & \begin{tabular}[c]{@{}l@{}}2. Which business category (e.g., grocery stores, full-service\\restaurants, limited-service restaurants, etc.)\\has the highest number of visits?\end{tabular}&         \checkmark & \checkmark  & x   & x  & \checkmark  \\  
\hline
\begin{tabular}[c]{@{}l@{}}Economic Management \\Public Health\end{tabular}                       & \begin{tabular}[c]{@{}l@{}}3. List the top ten bars in the state of Minnesota by \\long visit duration. A long visit duration is a visit that\\lasts longer than 20 minutes.\end{tabular}   &   \checkmark & \checkmark  & x & x  & \checkmark \\ 
\hline
Economic Management  & \begin{tabular}[c]{@{}l@{}}4. Which business category has the least impact from \\COVID-19 pandemic and related policy interventions \\since the stay-at-home order took effect in early~April?\end{tabular} & \checkmark & x & x & \checkmark & x \\
\hline
\begin{tabular}[c]{@{}l@{}}Public Transportation \\and Transit\end{tabular}                       & \begin{tabular}[c]{@{}l@{}}5. What is the difference in median distance traveled\\between commuters (e.g., people who go shopping)\\and delivery vehicles (e.g., USPS, DoorDash)?\end{tabular}               & x          & {\cellcolor[rgb]{0.396,0.396,0.396}}                                                       & {\cellcolor[rgb]{0.396,0.396,0.396}}                                       & {\cellcolor[rgb]{0.396,0.396,0.396}}                                          & {\cellcolor[rgb]{0.396,0.396,0.396}}                           \\ 
\hline
Public Health                                                                                     & \begin{tabular}[c]{@{}l@{}}6. Does Census Block Group have the highest number\\~of confirmed COVID-19 cases?\end{tabular}                                                                                    & x          & {\cellcolor[rgb]{0.396,0.396,0.396}}                                                       & {\cellcolor[rgb]{0.396,0.396,0.396}}                                       & {\cellcolor[rgb]{0.396,0.396,0.396}}                                          & {\cellcolor[rgb]{0.396,0.396,0.396}}                           \\ 
\hline
Traffic Flow   Public Safety                                                                      & \begin{tabular}[c]{@{}l@{}}7. How many smartphones reported from the \\I-35W bridge during a given protest event?\end{tabular}                                                                               & x          & {\cellcolor[rgb]{0.396,0.396,0.396}}                                                       & {\cellcolor[rgb]{0.396,0.396,0.396}}                                       & {\cellcolor[rgb]{0.396,0.396,0.396}}                                          & {\cellcolor[rgb]{0.396,0.396,0.396}}                           \\ 
\hline
Economic Management                                                                               & \begin{tabular}[c]{@{}l@{}}8. What is the difference in the unemployment rate\\based on gender, and which business category, and\\what brand contributes the most to the unemployment rate?\end{tabular}     & x          & {\cellcolor[rgb]{0.396,0.396,0.396}}                                                       & {\cellcolor[rgb]{0.396,0.396,0.396}}                                       & {\cellcolor[rgb]{0.396,0.396,0.396}}                                          & {\cellcolor[rgb]{0.396,0.396,0.396}}                           \\
\hline
\end{tabular}
\end{table}


The first four queries could be answered using the existing dataset since the required data, i.e., spatial extent of each census block group, the number of visits, Points Of Interests (POIs), and visit duration are available. On the other hand, the last four queries may provide useful and interesting patterns that can help policymakers; however, due to a lack of information (e.g., confirmed COVID-19 cases and unemployment rates) such queries could not be executed. To investigating these types of queries requires integrating complementary datasets. 

\subsection{Advantages of Relational Database Management System and Normalized Schema}
\label{advantageRDBMS}
Single flat database table design have two important limitations. The first is data redundancy which happens when attributes are duplicated because they need to be stored with other attributes on which they are not dependent. The second is design limitation of a flat table which limits the set of queries that can be easily answered using database query language SQL. This limits many frequently used queries that require summary statistics such as aggregation and ranking functions (e.g., AVERAGE, ORDER BY). For instance, certain composite data types (e.g., key-value pairs) are stored as a BLOB (binary large object) or CLOB (character large object) in a flat table. Such objects require additional tools for post-processing which may not be familiar to the end-user. Further, the flat table design typically operates on a single value column and does not extend well to multi-value data attributes. Table \ref{flat_table} shows a flat table example of a Weekly Patterns Schema with a subset of attributes including Bucket\_Dwell\_Time which stored a list of CLOB objects where each CLOB object is a key-value pair of Dwell\_Time and Visits.
Relational Database Management System (RDBMS) reduces such limitations in flat table design via \textbf{data normalization}. The rest of this section briefly reviews some advantages of RDBMS over flat table design:

\begin{table}\scriptsize
\centering
\caption{Flat Table for Subset of Attributes in Weekly Patterns Dataset \textcolor{blue}{(Best in Color)}}
\label{flat_table}
\begin{tabular}{|l|l|l|l|l|} 
\hline
safegraph\_place\_id & Start\_Date & End\_Date & NAICS\_Code & Bucketed\_Dwell\_Time\\ 
(String) & (String) & (String) & (Integer) & (CLOB)\\ 
\hline
ID1 & 03-01-2020 & 05-01-2021 & 722410 & [(DwellTime\_1:Visits\_1), ... , (DwellTime\_n:Visits\_n)]\\ 
\hline
\end{tabular}
\end{table}

\textbf{(1) Reducing Application Development Time:} SQL is a declarative language where queries specify "What" but not "How" (algorithms and data structures). Thus SQL queries take fewer lines of code than corresponding programs in procedural languages such as Python and Java. Consequently, SQL takes less time to design, develop and test applications than corresponding procedural languages. For instance, say we can produce a similar table of aggregated visits by Dwell\_Time in Figure \ref{bars} by executing a few lines of SQL code:\\
\newline
\textbf{Query 1:} List all aggregated visits based on Dwell Time for Bars from ‘03-01-2020’ to ‘06-28-2021’\\

\noindent \textbf{SELECT} DwellTime, SUM(Visits)\\
\noindent \textbf{FROM} Weekly Patterns \\
\noindent \textbf{WHERE} NAICS Code = ‘722410’ \\
\noindent \textbf{AND} (Start\_Date >= ‘03-01-2020’ \textbf{AND}  End\_Date <= ‘06-28-2021’) \\
\noindent \textbf{GROUP BY} DwellTime\\

\textbf{(2) Normal Form Table Design to Support Ad-hoc Query}: The design of relational tables and schema impact the ability and flexibility to support unanticipated future queries. Unnormalized table designs provide limited flexibility, mainly if they contain columns with multiple values or string encoding of complex data. The normal form.  table designs improve flexibility. For instance, the single-column containing Bucket\_Dwell\_Time can be split into two separate columns of Dwell\_Time and Visits along with their respective primitive data types as shown in Table \ref{1NF}. However, certain multivalued attributes (e.g., list in Bucket\_Dwell\_Time) can result in redundancies (e.g., as nested relations) \cite{navathe2001fundamentals}. For instance, Bucket\_Dwell\_Time shown in Table \ref{flat_table} contains multivalued attributes as a list of dictionary objects. Such redundancies are resolved by data normalization via the first normal form (1NF), which preserves atomicity within a table \cite{navathe2001fundamentals}. Table \ref{1NF} shows the normalized version ($1^{st}$ Normal Form) along with user-defined operations over Table \ref{flat_table} where each element (Dwell\_Time and Visits pair) in the multivalued row of Bucket\_Dwell\_Time is assigned to their respective "safegraph\_place\_id". Hence, it preserves the atomicity since no row has any multi-attributes. While this can also lead to partial and transitive dependencies, these can be resolved further by higher-order normal forms. \cite{navathe2001fundamentals}.

\begin{table}\scriptsize
\centering
\caption{ $1^{st}$ Normal Form Normalized Table for Subset of Attributes in Weekly Patterns Dataset \textcolor{blue}{(Best in Color)}}
\label{1NF}
\begin{tabular}{|l|l|l|l|l|l|} 
\hline
safegraph\_place\_id & Start\_Date & End\_Date & NAICS\_Code & Dwell\_Time & Visits\\ 
(String) & (String) & (String) & (Integer) & (String) & (Integer)\\ 
\hline
ID1 & 03-01-2020 & 03-08-2020 & 722410 & DwellTime\_1 & Visits\_1\\ 
\hline
ID2 & 03-08-2020 & 03-15-2020 & 722410 & DwellTime\_2 & Visits\_2\\ 
\hline
\end{tabular}
\end{table}

\textbf{(3) Compatibility with Application Programming Interface:} Application Programming Interfaces \cite{fielding2000architectural} (API) allow developers and third parties to access the data programmatically. Structured Query Language (SQL) itself is a standardized API that is very portable and can further be extended to other programming languages via Open Database Connectivity (ODBC) and its wrappers such as Java Database Connectivity (JDBC) and Pysql. The use of API has several benefits. For example, developers do not need to know how an API is implemented or the underlying data structure and schema. Hence, it serves the purpose of data security and minimal code requirements on the data-user side. There are various API protocols (e.g., SOAP, XML-RPC, REST, etc.). Other details can be found here \cite{APITypes}. REST APIs can be used to access data similar to that in Figure \ref{bars} by which internally executing Query 1 and returning the response in various formats (e.g., JSON, XML). For example, a REST API call for Query 1 may be as follows:\\

\url{<server-address>/Visits?Code=722410&Start_Date>=`03-01-2020'&End_Date<=`06-28-2021'}\\

where, server-address is the address of the server hosting the API and ”/Visits” is the handle to address queries regarding the total number of visits by dwell time between a start date and an end date for a given point of interest (POI) identified by its NAICS code. Relational databases in 3rd normal form (3NF) make it possible to design simpler APIs which require fewer parameters to access the required information.

The use of APIs can help expand the usability of a dataset across different jurisdictions (e.g., city, county, state) at different scales (city-wide, county-wide, etc.). For example, separate handles can be provided to two different counties to access the desired information.

Further, several design decisions at the database level can be taken to improve query efficiency. For example, databases can be designed to limit data merges across different locations for efficiency. Most importantly, domain officials and other stakeholders are not required to know all these details to use the API calls.

\textbf{(4) Query Processing and Optimization}: Query processing and optimization refer to choosing an efficient query execution plan based on the existing storage structures. The decisions about which indexes to create and maintain are part of the physical database design of a relational database, and tuning is based on the given table properties (e.g., table size, etc.). For instance, performing a linear search in a small table proves to be more efficient than using an index-based search.

Other advantages of RDBMS include centralized environments and security and availability (e.g., fault tolerance, recovery, etc.) to all user groups.

\section{Validation}{\label{sec:Validation}}

\begin{figure*}[!ht]
    \centering
    \includegraphics[width=1.0\textwidth]{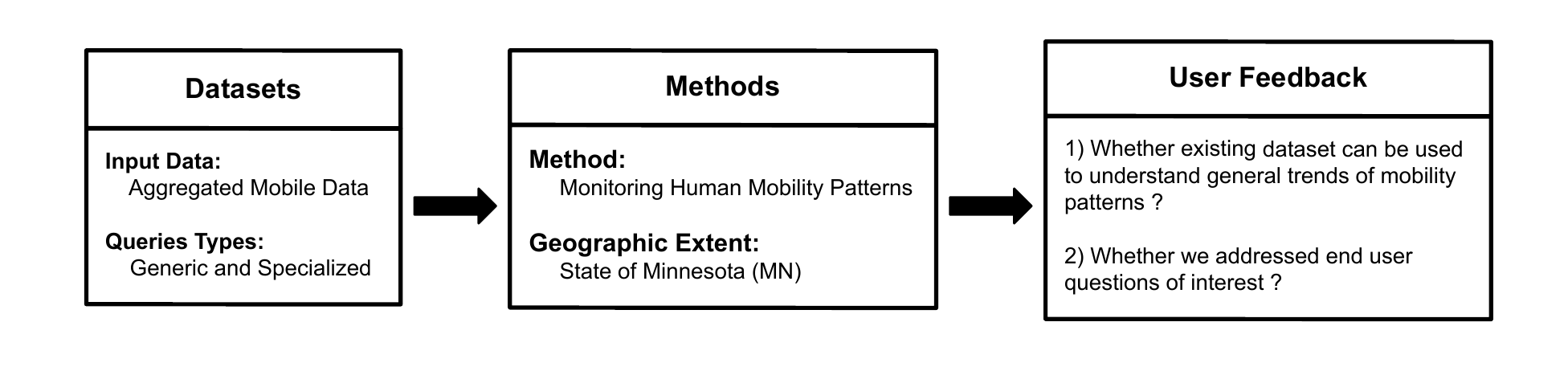}
    \caption{Validation Framework}
    \label{validation_framework}
\end{figure*}

We validated the proposed decision support system via a case study using a real world human mobility dataset for Minnesota. Section \ref{dataset} describes SafeGraph COVID-19 Data Consortium \cite{safeGraph2020} data, including Weekly Patterns and Social Distancing Dataset. In addition, we briefly talk about the Minnesota calendar of policy interventions. In Section \ref{Map_val}, we investigate weekly visit trends via a case-study over top business categories with long-duration visits greater than 20 minutes. We further explore frequently of visits in various categories at different bucket dwell times, which leads us to find new interesting spatiotemporal patterns. The overall validation framework is shown in Figure \ref{validation_framework}. 
\subsection{Dataset Description and Case Study in Minnesota}\label{dataset}
The dataset description and Case Study in Minnesota as described in Figure \ref{validation_framework} and MN Policy Calendar (shown in Table \ref{calender_table}) are as follows:

\textbf{SafeGraph:} The mobility data in this work was supported by COVID-19 Response SafeGraph Data Products \cite{safeGraph2020}. The SafeGraph COVID-19 Data Consortium is providing various free datasets (e.g., social distancing, weekly patterns, monthly patterns, etc.) to academic researchers, non-profit organizations, and governments to study the primary and secondary effects of the novel coronavirus. The raw data was generated by using a panel of positioning system pings from anonymous mobile devices. We used weekly pattern data to monitor mobility across the most frequently visited business categories. The current weekly pattern data is given in a single flat table format (see Figure \ref{Denormalized}) that consists of many data attribute columns with a short description of each attribute. A core place table provides information such as a North American Industry Classification System (NAICS) \cite{naics2021} categorizing business categories for Points Of Interests, which used in conjunction with Weekly Patterns. Social Distancing Metrics provide aggregated statistics of distance traveled within each Census Block Group.

\textbf{A Case Study in Minnesota: }As the case study in this work was conducted in Minnesota, we describe the summary statistics of SafeGraph in this context. The Safegraph data for Minnesota was derived from 294,014 individual mobile devices, which is roughly 5\% of the total state population. Further, it covers 73,548 points of interest (POIs) across 261 different business categories (e.g., full-service restaurants, limited-service restaurants, gasoline stations with convenience stores). Due to privacy concerns and data protection, SafeGraph aggregates individual-level anonymous data points to the census block group (CBG). Its spatial coverage in Minnesota is 4107 out of 4111 \cite{census2020} CBGs. This data was used for both specific and generic query types. For validating our community-based decision support system, we provide mobility analysis for specific queries related to long visits duration.

\textbf{MN Policy Intervention Calendar:} After figuring out which types of mobility questions were answerable with the available data, we then analyzed mobility patterns based on MN Policy Intervention Calendar, as shown in Table \ref{calender_table}. These interventions were imposed by the state government in order to control and minimize the spread of COVID-19. The policy intervention calendar starts on March 17, 2020, with the University of Minnesota (U of M) School Closing ordered just after Spring Break (i.e., March 9, 2020).  This order was followed by the Minnesota Stay-at-Home order issued on March 27, 2020. After holding a Stay-at-Home order for a month, a small reopening phase was ordered with gatherings of 10 or less. Subsequent re-openings, namely, Phase 2 and Phase 3, were issued on June 1 and June 10, 2020, respectively, along with gradual increasing of gathering capacity for essential services. Due to an increase in COVID incidents, the state government issued another shutdown for certain business categories in mid-November which later reopened in early January. Further restrictions related to capacity and social distancing were lifted on May 27, 2021.
\begin{table}\scriptsize
\centering
\caption{MN COVID-19 policy Interventions Calendar \cite{calendar2020} \textcolor{blue}{(Best in Color)}}
\label{calender_table}
\begin{tabular}{|l|l|} 
\hline
Dates         & Social setting                          \\ 
\hline
Mar 9, 2020   & University of Minnesota Spring break    \\ 
\hline
Mar 17, 2020  & University of Minnesota school closing  \\ 
\hline
Mar 27, 2020  & MN stay-at-home                         \\ 
\hline
May 18, 2020  & MN reopening Phase 1                    \\ 
\hline
June 1, 2020  & MN reopening Phase 2                    \\ 
\hline
June 10, 2020 & MN reopening Phase 3                    \\
\hline
November 16, 2020 & MN shutdown order for Bars and Restaurants \\
\hline
January 11, 2021 & MN reopening order for Bars and Restaurants \\
\hline
May 27, 2021 & No limits on size and no social distancing requirements. \\
\hline
\end{tabular}
\end{table}
\subsection{Mapping and analyzing human mobility changes}\label{Map_val}
To validate our schema with end-user queries, we provide a mobility analysis based on number of long-duration visits (i.e. visits with durations greater than 20 minutes) aggregated by weeks for about one year (i.e. from March 1, 2020 to June 28, 2021).

\begin{figure*}[!ht]
    \centering
    \includegraphics[width=1.0\textwidth]{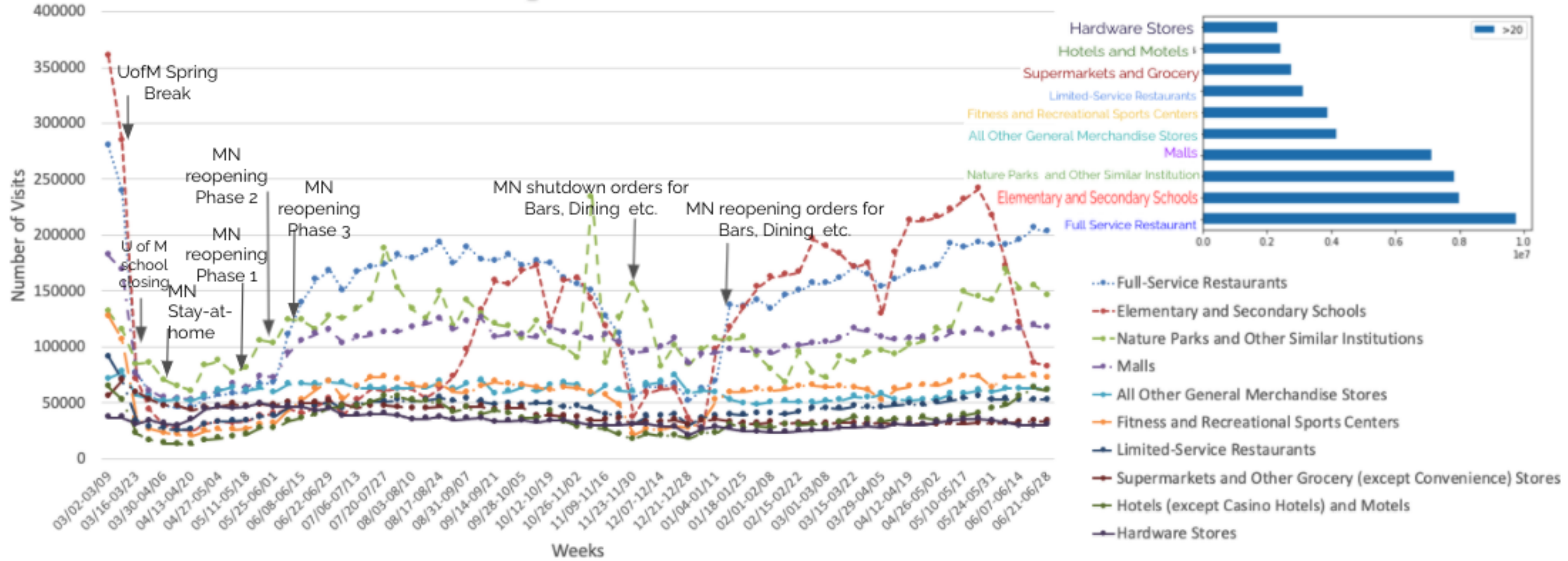}
    \caption{Most frequent long-duration visited business categories in Minnesota (March 2 ,2020 - June 28, 2021) \textcolor{blue}{(Best in Color)}.}
    \label{buisness_categories}
\end{figure*}

We first created summary reports of the most frequently visited business categories (as shown in Fig. \ref{buisness_categories}) based on long-duration visits. This led us to distinguish anomalous behaviors and mobility impacts in different business categories. For example, elementary and secondary schools were the most frequented business categories in early March 2020. But starting at 2020 spring break, and shortly after the closure of the schools, we observed an enormous drop in that category, and it has stayed low over the course of the pandemic. On the other hand, full-service restaurants, malls, and natural parks had an extensive initial drop; however, subsequently they illustrate positive trends, and almost recover to their original levels even before the start of the first phase of the opening. Next, we evaluated them based on the duration of people's visits.


Using the ERD, we quantified interactions based on bucket dwell times (e.g., visits longer than 20 minutes). We produced a summary of hangout place reports that identified POIs with long-duration visits. Classifying the frequency of visits based on visit duration allowed us to identify business categories (e.g., full-service restaurants, bars), which despite having relatively fewer visits in comparison with general categories (e.g., grocery stores), included more hotspots of disease \cite{bars2020a, bars2020b}. 


\begin{figure*}[!ht]
    \centering
    \includegraphics[width=1.0\textwidth]{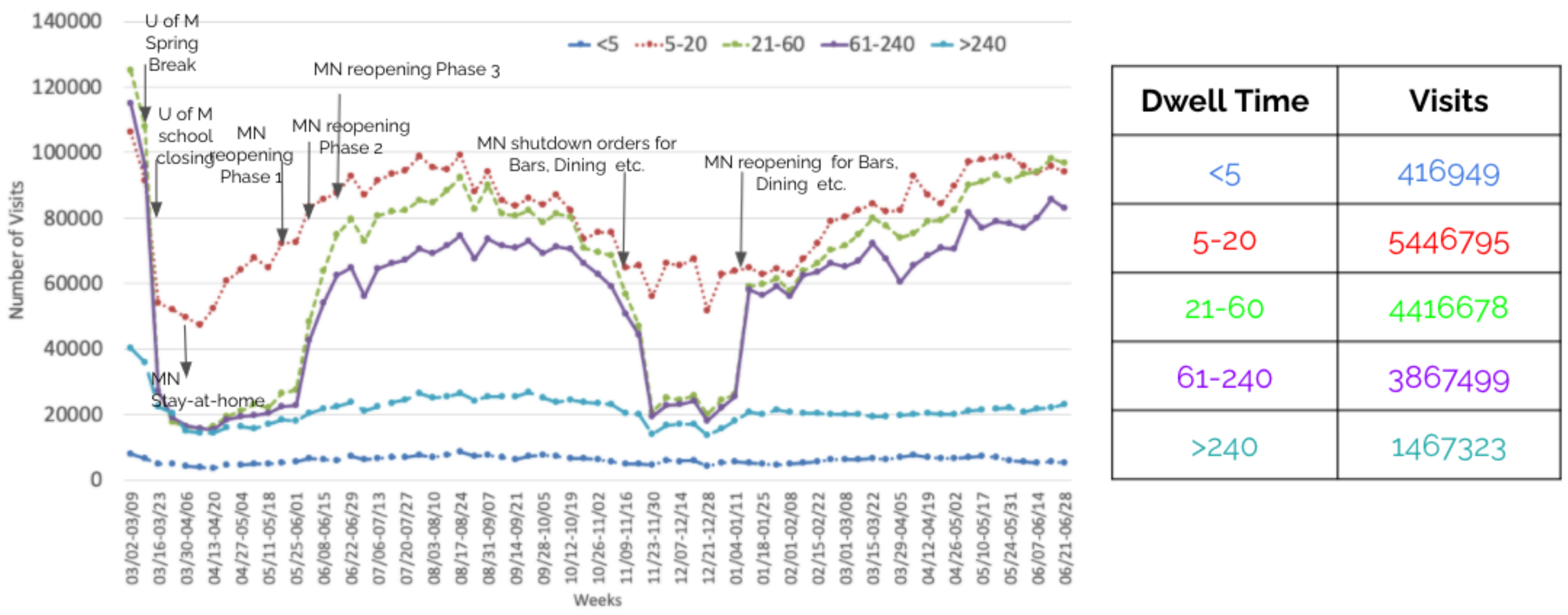}
    \caption{Analysis of full-service restaurants visits based on the bucket dwell times \textcolor{blue}{(Best in Color)}.}
    \label{Full-Service-Restaurants}
\end{figure*} 

\begin{figure*}[!ht]
    \centering
    \includegraphics[width=1.0\textwidth]{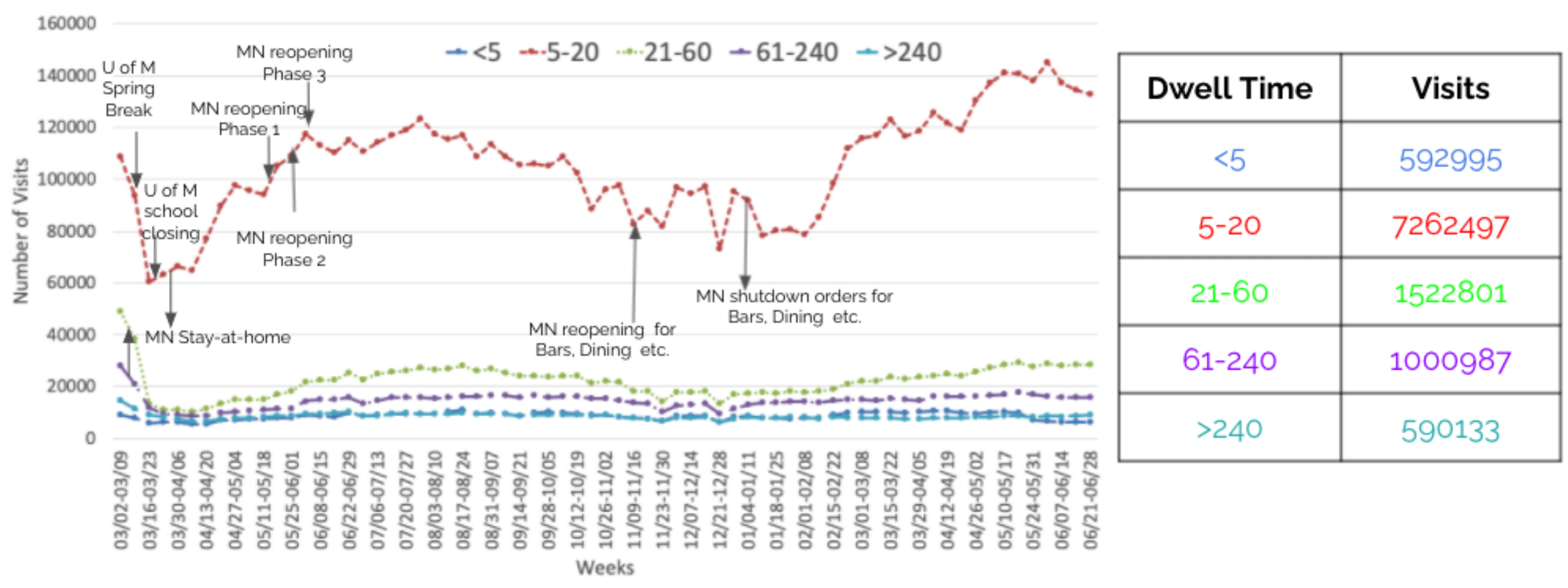}
    \caption{Analysis of Limited Service Restaurants based on the bucket dwell times \textcolor{blue}{(Best in Color)}.}
    \label{Limited-Service-Restaurants}
\end{figure*}

\begin{figure*}[!ht]
    \centering
    \includegraphics[width=1.0\textwidth]{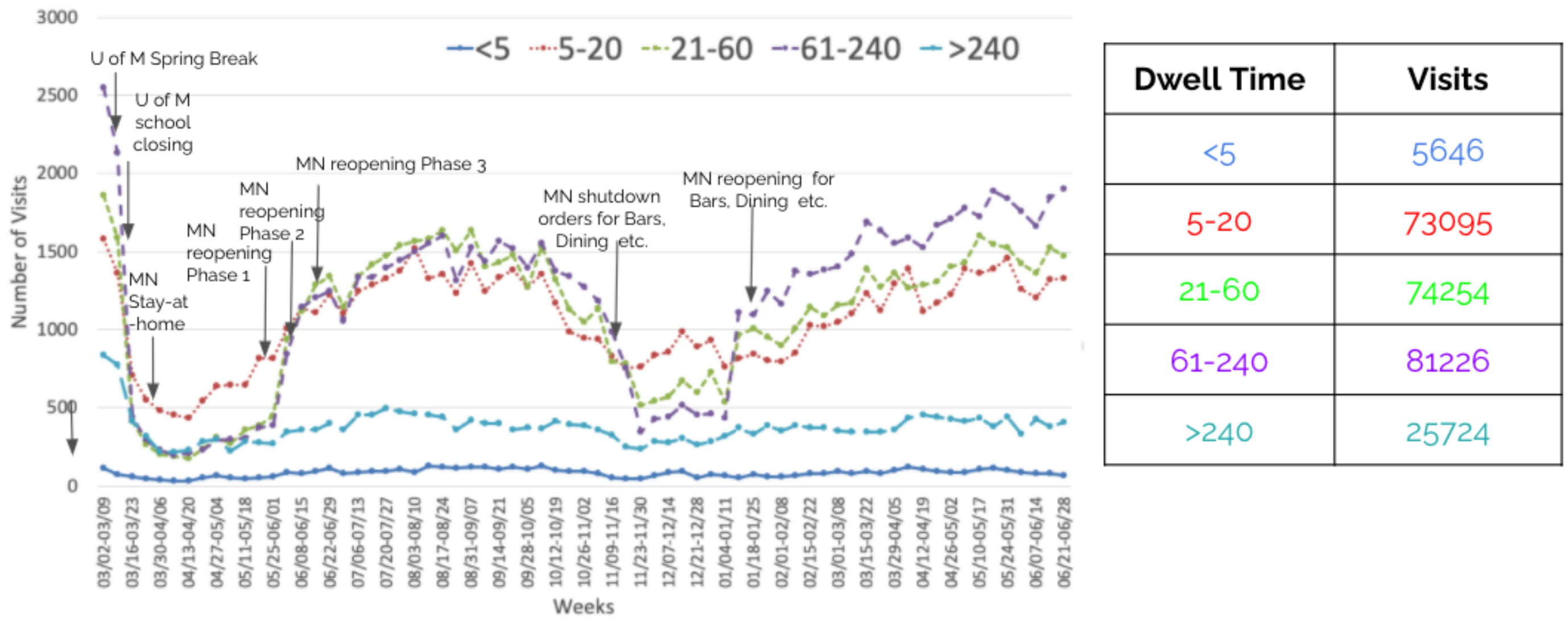}
    \caption{Analysis of bar visits based on bucket dwell times \textcolor{blue}{(Best in Color)}.}
    \label{bars}
\end{figure*}

Figures \ref{Full-Service-Restaurants}, \ref{Limited-Service-Restaurants} and \ref{bars} show the frequency of visits to different business categories, including full-service restaurants, limited-service restaurants, and bars, based on various bucket dwell times ranging from shorter visits to longer visits (e.g., 61-240 minutes). We observed a rapid increase to bars and full-service restaurants that emerged shortly by the first phase of reopening (Figures \ref{Full-Service-Restaurants} and \ref{bars}), whereas the same is not true in case of Limited-Service Restaurants (Figure \ref{Limited-Service-Restaurants}).

We also compared these trends with the outbreak reports by the Minnesota Department of Health (MDH) \cite{bars2020d} that specifically name bars and restaurants deemed as venues with significant impacts on COVID spread \cite{bars2020c} \cite{bars2020f} that were linked to COVID-19 cases during different months.
According to the MDH \cite{bars2020g}, an outbreak location is a restaurant or bar with at least seven unrelated COVID cases from seven different households that only visited one restaurant or bar establishment during that month. To find patterns that exclusively apply to outbreak locations, we included non-outbreak locations to act as a control group which was selected to keep multiple external variables between the two groups constant. For each matched pair of an outbreak and non-outbreak location, we picked places within proximity of each other to maintain consistent local mobility patterns for both locations. Then, we ensured that locations were both part of the same business category either both full-service restaurants or both bars. Lastly, we checked that both locations had a similar number of weekly visits to their locations before the most recent COVID-19 lockdown: either visits before the March or November shutdown of indoor dining, depending on the month being analyzed.
\begin{figure*}[!ht]
    \centering
    \includegraphics[width=0.6\textwidth]{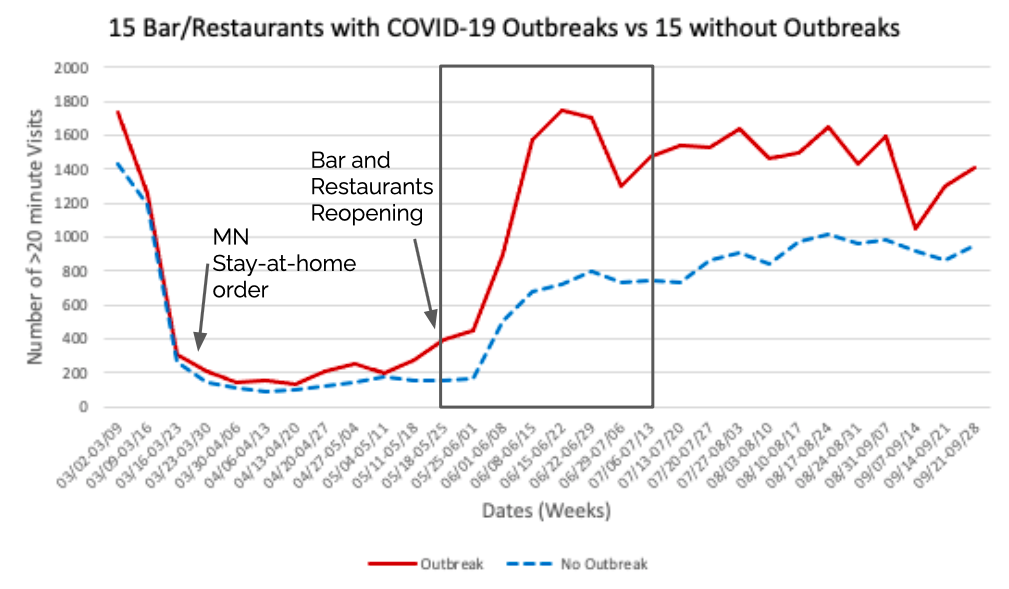}
    \caption{Long-duration visits for outbreak and non-outbreak groups \textcolor{blue}{(Best in Color)}.}
    \label{outbreak}
\end{figure*}

We analyzed the trend of long-duration visits to 15 locations linked to COVID outbreaks in June 2020 and 15 non-outbreak locations. As shown in Figure \ref{outbreak}, we first notice an increase in long-duration visits for both outbreak and non-outbreak groups after the reopening period, which is expected. However, by June 29, 2020, the outbreak group’s long-duration visits reached pre-COVID levels of visits (first week of March), whereas the no-outbreak group only reached 50\%. That suggests that large increases in long-duration visits are associated with outbreaks. However, such observations needs to be further validated by disease transmission model (e.g., SIR/SEIR \cite{martcheva2015introduction}) simulations which consider indoor information, such as mask use and spatial distance between individuals (e.g., where they are sitting, etc.).

\section{Closing the Loop: User Feedback}\label{user_feedback}
To assess how well our decision-making system met the needs of the users, we sought feedback from our collaborators in transportation, economic management and public health. We begin by discussing their concerns with data quality, then we summarize other feedback such as the value derived and the need for additional information to interpret reports. This section provides a summary of the feedback.


\subsection{Data Quality}\label{dataquality}
In Spring 2020, many end-users expressed concerns about Safegraph data such as the following: 
\begin{itemize}
    \item How does the dataset protect the privacy of individuals ?
    \item What is the sampling bias ?
    \item Is the spatial and temporal granularity reasonable for the sample size ?
    \item How does the mobile device data compare with well known datasets in our domain ?
\end{itemize}
Next, we share our findings in context of the above questions.

\subsubsection{How does the dataset protect the privacy of individuals ? }\label{subsubsec:privacy}
\begin{figure*}[!ht]
     \begin{subfigure}[b]{0.30\textwidth}
          \centering
          \includegraphics[width=40mm, height=40mm]{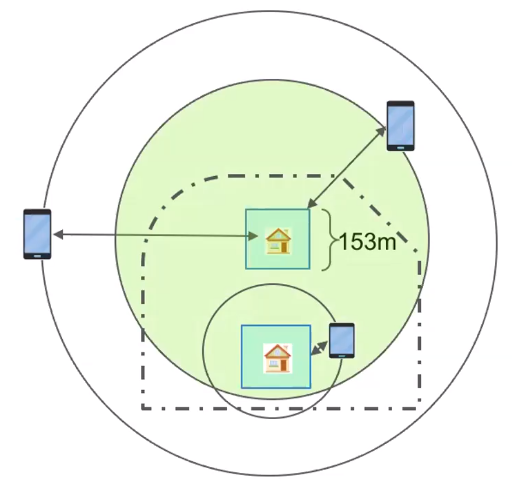}
          \caption{An illustrative example in reporting median range for a Census Block Group \textcolor{blue}{(Best in Color)}.}
          \label{privacy}
     \end{subfigure}
     \begin{subfigure}[b]{0.50\textwidth}
        \centering
        \includegraphics[width=60mm, height=50mm]{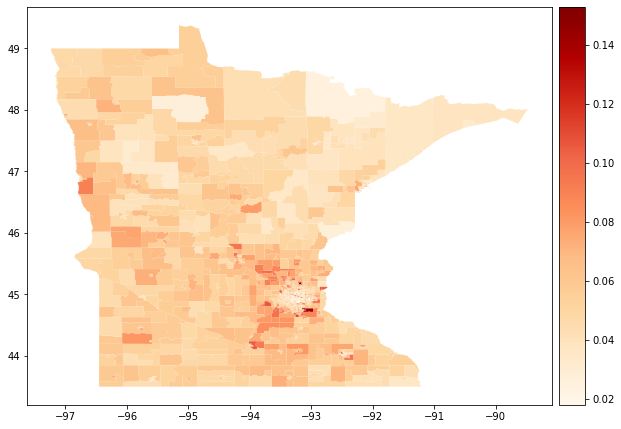}
        \caption{A map of county-level sampling rate in Minnesota.}
        \label{fig:state_device_ratio_wo}
     \end{subfigure}
        \caption{Data Quality assessment based on (a) data privacy and (b) sampling bias \textcolor{blue}{(Best in Color)}.}
        \label{Metropolitan}
\end{figure*}

The mobile device data aggregates to census block groups (CBGs) because a geo-referencing for a home's location includes a $153m\times153m$ buffer around the home, as shown in Fig. \ref{privacy}. However, this geocoding referencing may be considered as diluted, as it underestimates the effect of mobile people such as college students. To apply additional privacy and data protection, the median in some data attributes (e.g., distance\_traveled\_from\_home, median\_home\_dwell\_time) are reported. 

For example, as shown in Fig. \ref{privacy}, three mobile devices are within the range of a census block group, and the median range from those three devices is reported as the traveled distance from home. It is worth mentioning that if a census block group has fewer devices, that census block group is suppressed. Lastly, similar to US Census 2020 data, noise has been applied to safeguard data for protecting individual information. 

\subsubsection{What is the sampling rate? How does it vary over the counties? Does vendor report any other sample bias (e.g., demographic)?}\label{subsubsec:sampling}
The visualization in Figure \ref{fig:state_device_ratio_wo} provides a sampling rate based on the  the ratio between sampled devices and population in each census tract in Minnesota. As can be seen, the sampling rate ranges from 1\% to 15\% and is unevenly distributed. This helps us to understand the sampling bias based on population distribution i.e., around suburban regions near the Minneapolis Downtown region. However, our sampling bias discussion is limited to population distribution due to the paucity of demographic information in the given Safegraph dataset. Sampling bias issues in Safegraph data were also raised in literature \cite{coston2021leveraging} such as lack of demographic interpretation. Because of the low sampling rate and uneven distribution, the data needs to be aggregated to a certain spatial resolution to avoid conveying misleading information. The original dataset groups the sampled mobile devices by the census block groups where their ``home''s are. However, there are census block groups with very few or no sampled devices, which likely does not represent the population within the areas.



\subsubsection{Is the spatial and temporal granularity reasonable for the sample size ? }\label{subsubsec:granularity}

The mobile device data is spatially aggregated at a variety of levels starting from the census block group level up to the state level for all fifty states. In terms of temporal granularity, each mobile device is tracked at intervals ranging from monthly to hourly scale. However, the significance of a particular granularity varies from one disciplinary field to another. For instance, hourly monitoring of infection provide more value to the public health domain than other domains. Further, US Census data provide privacy for protecting individual information result in low accuracy as mentioned in Section \ref{subsubsec:privacy}. Such spatial aggregations is the most crucial factor affecting the quality of the data.

\subsubsection{How does the mobile device data compare with well known datasets in our domain ? }\label{subsubsec:comparison}

Due to the data quality issues within the existing dataset, we needed to know whether the trends found in the mobile device dataset could be validated by ground truth, such as loop detector data. To validate, we visualized the vehicle miles traveled using loop detector data from the Minnesota Metropolitan Council \cite{loop_detectors}. The dataset consists of observed vehicle miles traveled and predicted vehicle miles traveled. Predicted values are calculated based on a generalized additive model \cite{hastie1990generalized} using historical data from previous years (January 2018 - Early March 2020). 

\begin{figure*}[!ht]
     \centering
     \begin{subfigure}[b]{0.30\textwidth}
         \centering
         \includegraphics[width=5cm, height=4cm]{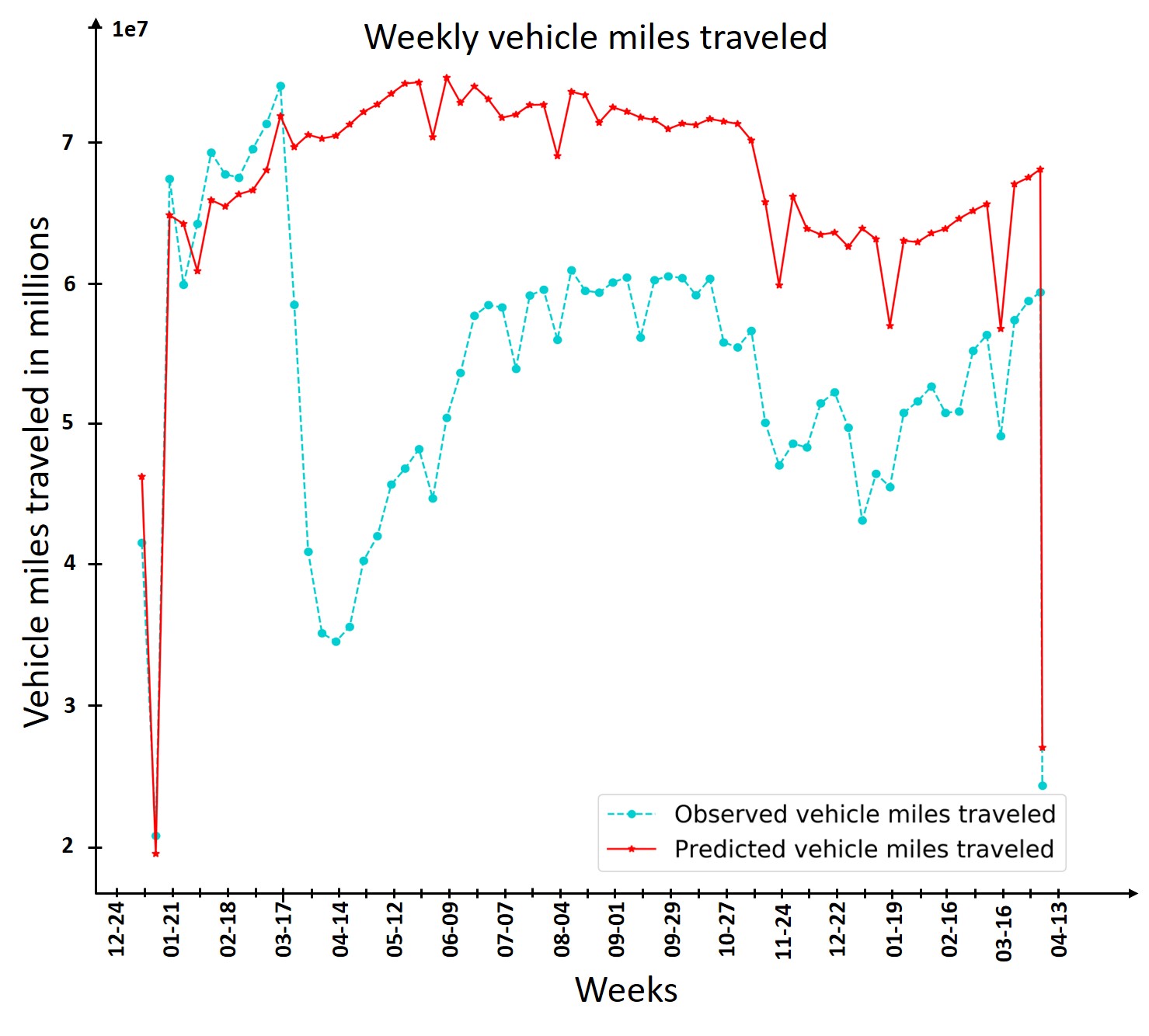}
         \caption{Predicted traffic volume}\label{pred_vs_observe}
     \end{subfigure}
     \begin{subfigure}[b]{0.50\textwidth}
         \centering
         \includegraphics[width=5cm, height=4cm]{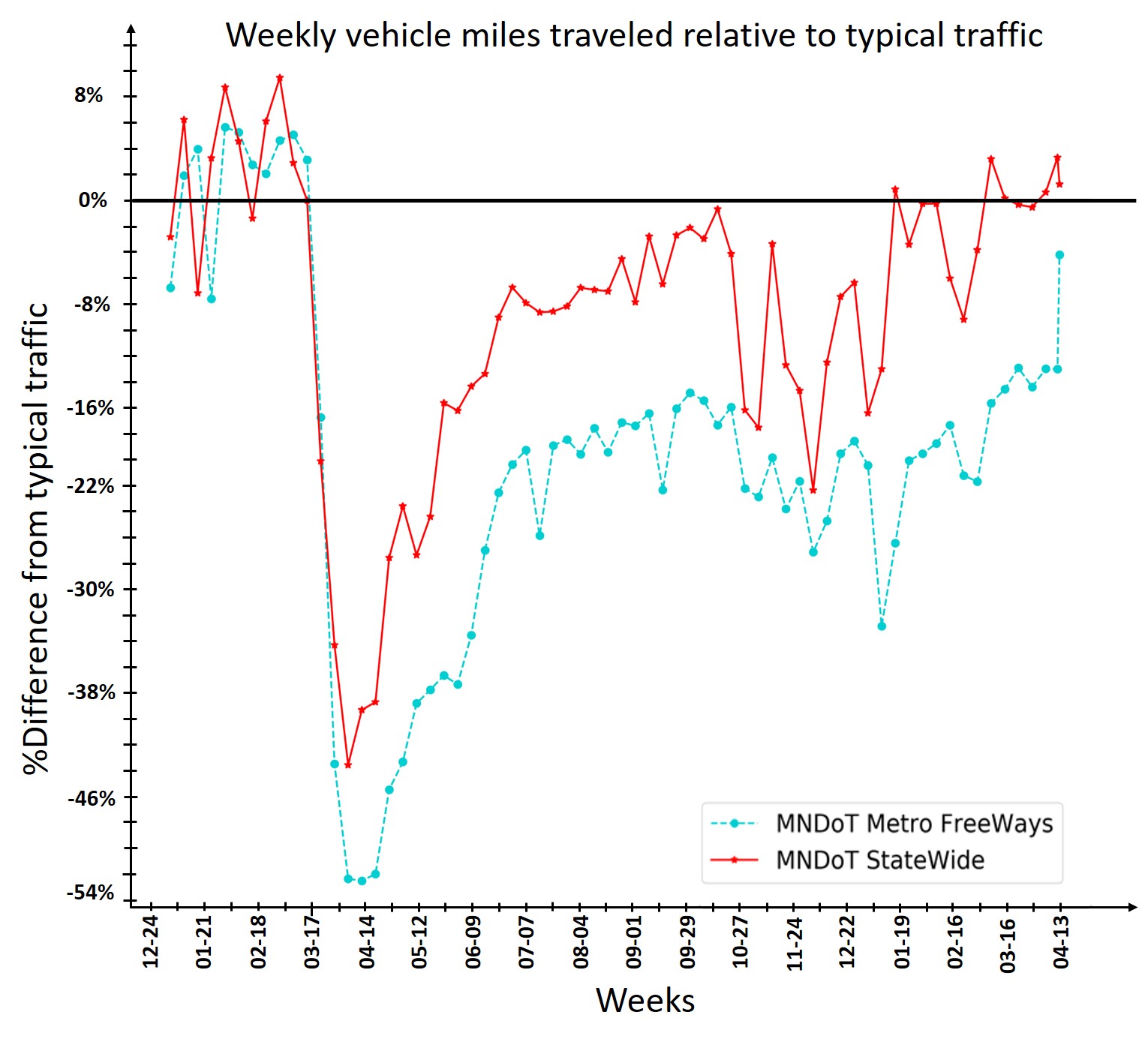}
         \caption{Traffic trends}\label{relatives}
     \end{subfigure}
        \caption{Decrease in traffic network across the COVID-19 timeline \textcolor{blue}{(Best in Color)}.}
        \label{Metropolitan}
\end{figure*}

Fig. \ref{pred_vs_observe} shows the trends for observed ad predicted miles traveled values over the entire freeway network in the year 2020. As can be seen, the network traffic flow dramatically dropped starting the third week of March compared to typical traffic with no outbreak or stay-at-home order in place. Fig. \ref{relatives} shows the relative trends in mobility for Minnesota in 2020, where the horizontal line at zero is the baseline, and the further below that line, the more traffic has decreased. As can be seen, the lowest point of network traffic for both the metro area freeway and statewide sensors happened just as the "stay-at-home" order took effect in late March.

\begin{figure*}[!ht]
     \centering
     \begin{subfigure}[b]{0.40\textwidth}
         \centering
         \includegraphics[width=5cm, height=5cm]{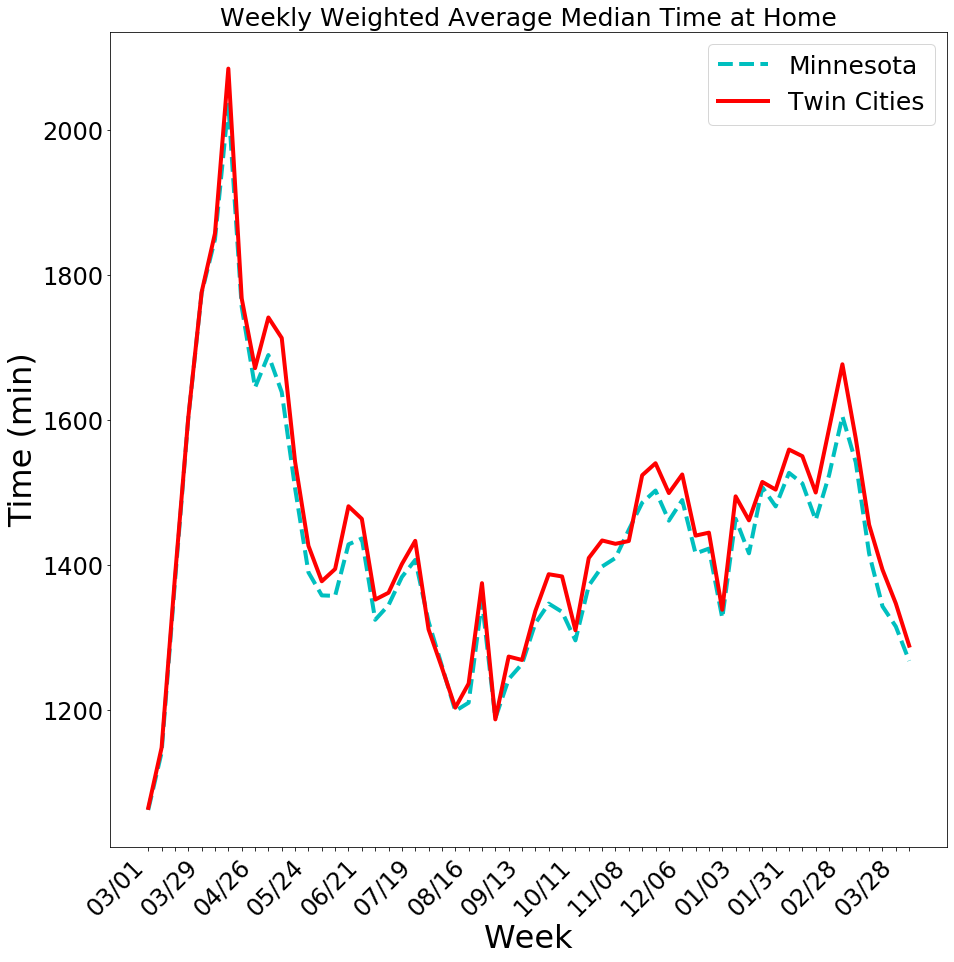}
         \caption{}
         \label{sl_tsh}
     \end{subfigure}
     \begin{subfigure}[b]{0.40\textwidth}
         \centering
         \includegraphics[width=5cm, height=5cm]{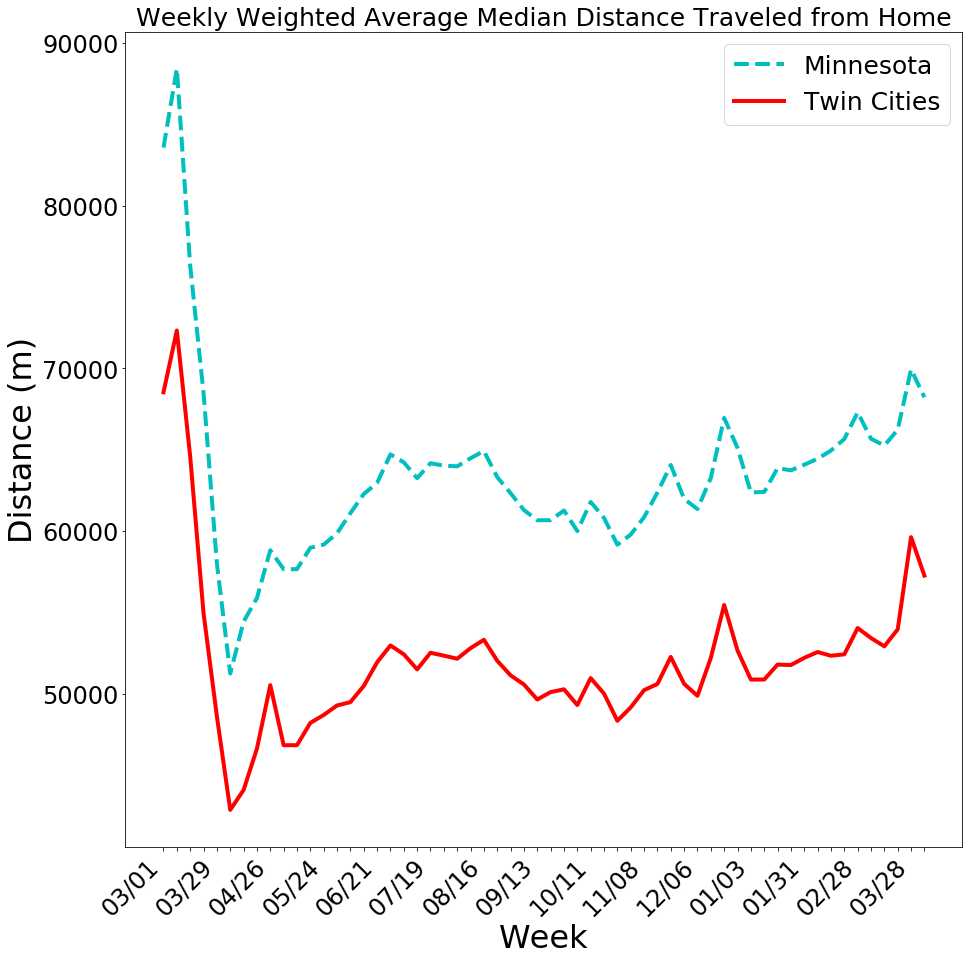}
         \caption{}
         \label{tc_tsh}
     \end{subfigure}
        \caption{(a) Weekly average time spent at home, (b) Weekly average distance travel from home\textcolor{blue}{ (Best in Color)}.}
        \label{confirm_safeGraph_SD}
\end{figure*}

Similarly, we created mobility reports on two key data attributes, ``time spent at home'' and ``distance travel from home'' using the social distancing dataset. As shown in Fig. \ref{confirm_safeGraph_SD}, we observed similar trends to the loop detectors dataset for both metropolitan areas as well as for the state of Minnesota. Given these comparably equivalent trends and despite the existing data quality issues, the mobile device dataset may be used to identify novel spatiotemporal patterns and further evaluate and understand the COVID-19 impacts on mobility at a fine geographic resolution.

\subsection{Information Value and Report Quality}\label{discussion}

\subsubsection{Information Value :}We sought feedback from users in our partner communities regarding the value of our reports to their decision making analysis. The feedback was obtained through multiple meetings, interview and email exchanges. The rest of this section summarizes the feedback by user groups, namely, Traffic Flow and Public Safety, Public Transportation and Transit, Economic Management and Public Health.

\begin{figure}[!ht]
    \centering
    \includegraphics[height=35mm]{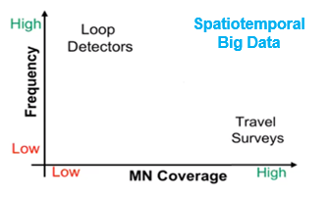}
    \caption{Frequency and geographic coverage of traditional datasets and spatial big data \textcolor{blue}{(Best in Color)}.}
    \label{STBD}
\end{figure}

\textbf{Traffic Flow and Public Safety:} Transportation analysts appreciated the capabilities of the mobile device spatiotemporal data based on geographic extent and sampling frequency compared with Loop Detectors and Travel Surveys respectively. The use of spatial big data in transportation research  \cite{shekhar2012spatial} is promising due to both high frequency and high geographic coverage (Fig. \ref{STBD}). 

Travel surveys are limited in number, and while loop detectors cover urban highways, they cannot capture public squares, avenues, parkways, etc. Further, analysts also acknowledged the potential of Safegraph to handle large mobile device data at certain geographic levels (e.g. census tracts, census block groups) irrespective of data quality issues. In addition, we also tried to address specific queries of interest. For instance, the transportation analysts were interested in separating vehicle miles traveled from commuting from vehicle miles traveled for delivery goods. However, we were unable to provide the results since it was not possible to slice vehicle miles traveled by mode of transportation (e.g. cars vs delivery trucks).

\begin{figure*}
    \centering
    \includegraphics[width=1.0\textwidth]{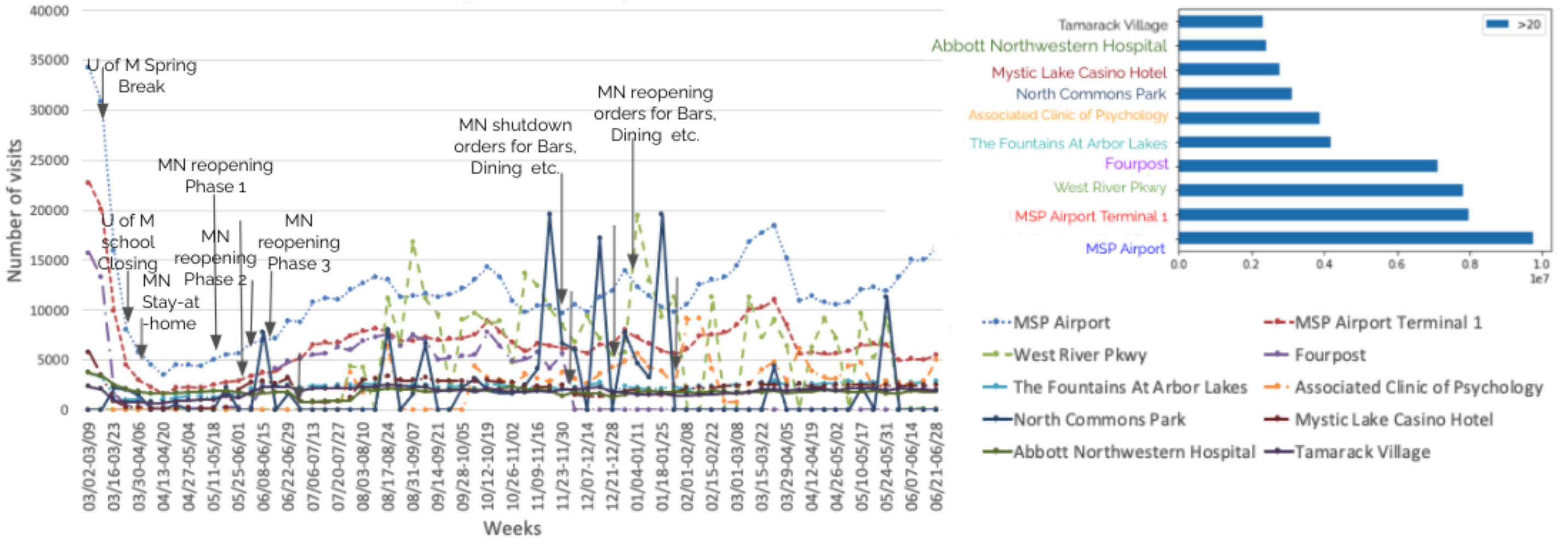}
    \caption{Most frequent long-duration visited Points of Interest in Minnesota (Since March 2, 2020 - June 28, 2021) \textcolor{blue}{(Best in Color)}.}
    \label{top_poi}
\end{figure*}

\textbf{Public Transportation and Transit:} Metro transit analysts acknowledged they found the weekly reports helpful for tracking the types of trips that might be recurring, and perhaps eventually returning to transit. They were interested in certain Points of Interest such as "Fourpost" (as shown in Figure \ref{top_poi}) since the number of visits there increased from 06/30/2020 but eventually went down to zero after 11/16/2020. According to a news report \cite{Fourpost2020d}, "Fourpost" was closed early 2020 before the stay-at-home order and offered physical space to other retailers, explaining some activity related to long or short duration visits after 06/30/2020. Hence, data quality issues still persist since the Safegraph dataset may not have been updating or removing old POIs frequently enough.

\textbf{Economic Management:} Economic management colleagues acknowledged the value of our human mobility patterns based on long visit duration with COVID-19 policy Interventions Calendar i.e., state's Stay-at-Home order as well as safely reopening the economy. They encouraged us to share weekly reports based on mobility traffic for certain places, businesses and business categories. We also conducted regular meetings to discuss questions posed by policymakers and analysts for certain business categories. However, the most significant patterns of interest to policymakers remained Bars, Limited and Full-time Restaurants.

\textbf{Public Health}: When discussing Safegraph data, public health researchers posed initial questions regarding data quality such as sampling bias and data transparency. However, such issues are not addressed in Safegraph data which presents a barrier to perform calibration in disease transmission models (SEIR). Nevertheless, they were also interested in ability to identify visits of long duration for estimating number of contacts with hourly granularity for each long visits.

\subsubsection{Report Quality:} We published periodic reports to Economic Management policymakers and analysts, and they found the report quality acceptable. In addition, they requested supplement at materials such as summary data in tabular format to supplement the trend via visualizations.\\

\begin{figure*}[!ht]
     \begin{subfigure}[b]{0.43\textwidth}
          \centering
          \includegraphics[width=50mm, height=50mm]{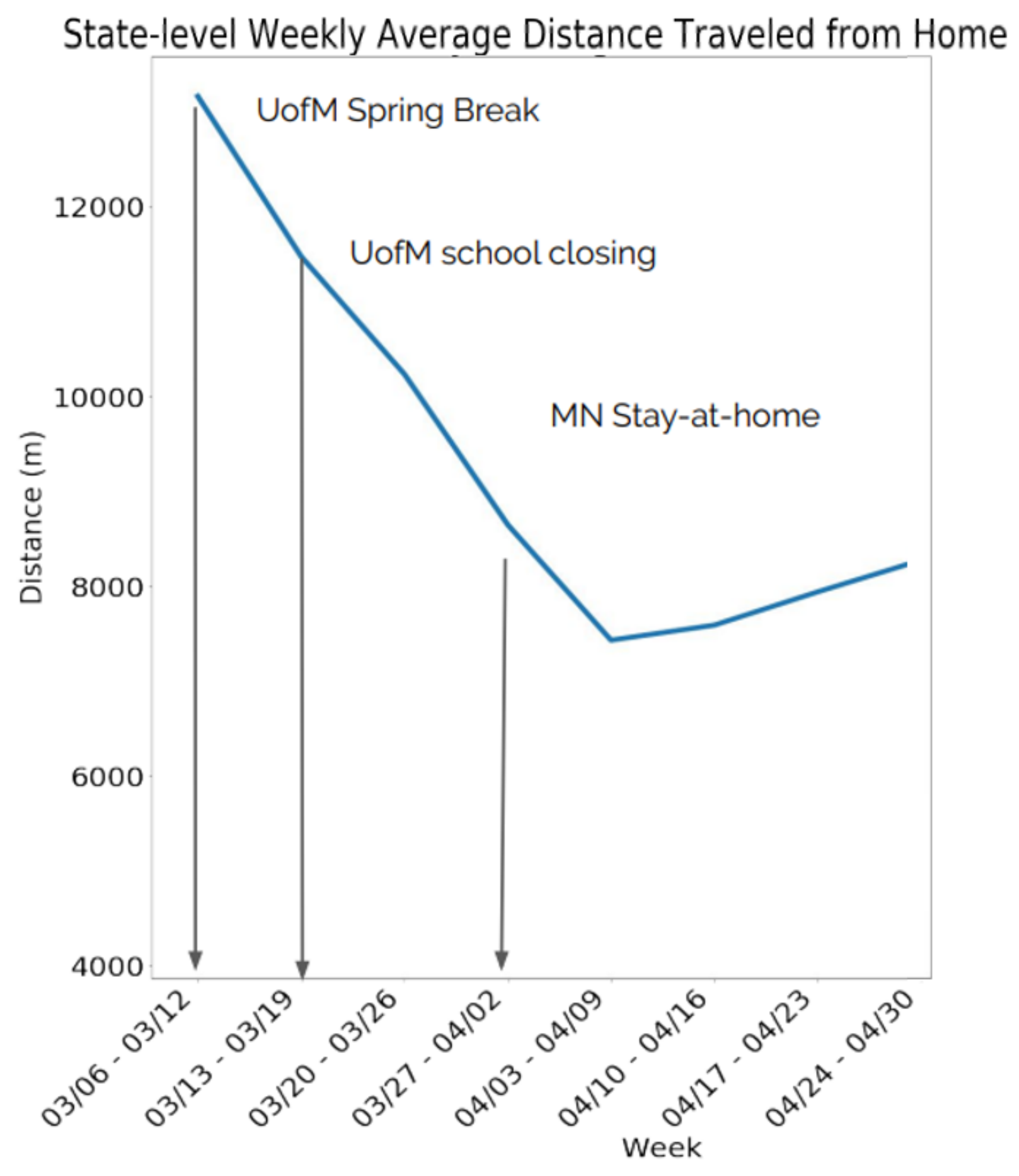}
          \caption{Average Distance Travelled from Home}
          \label{fig:avg_dist_from_home}
     \end{subfigure}
     \begin{subfigure}[b]{0.41\textwidth}
        \centering
        \includegraphics[width=50mm, height=50mm]{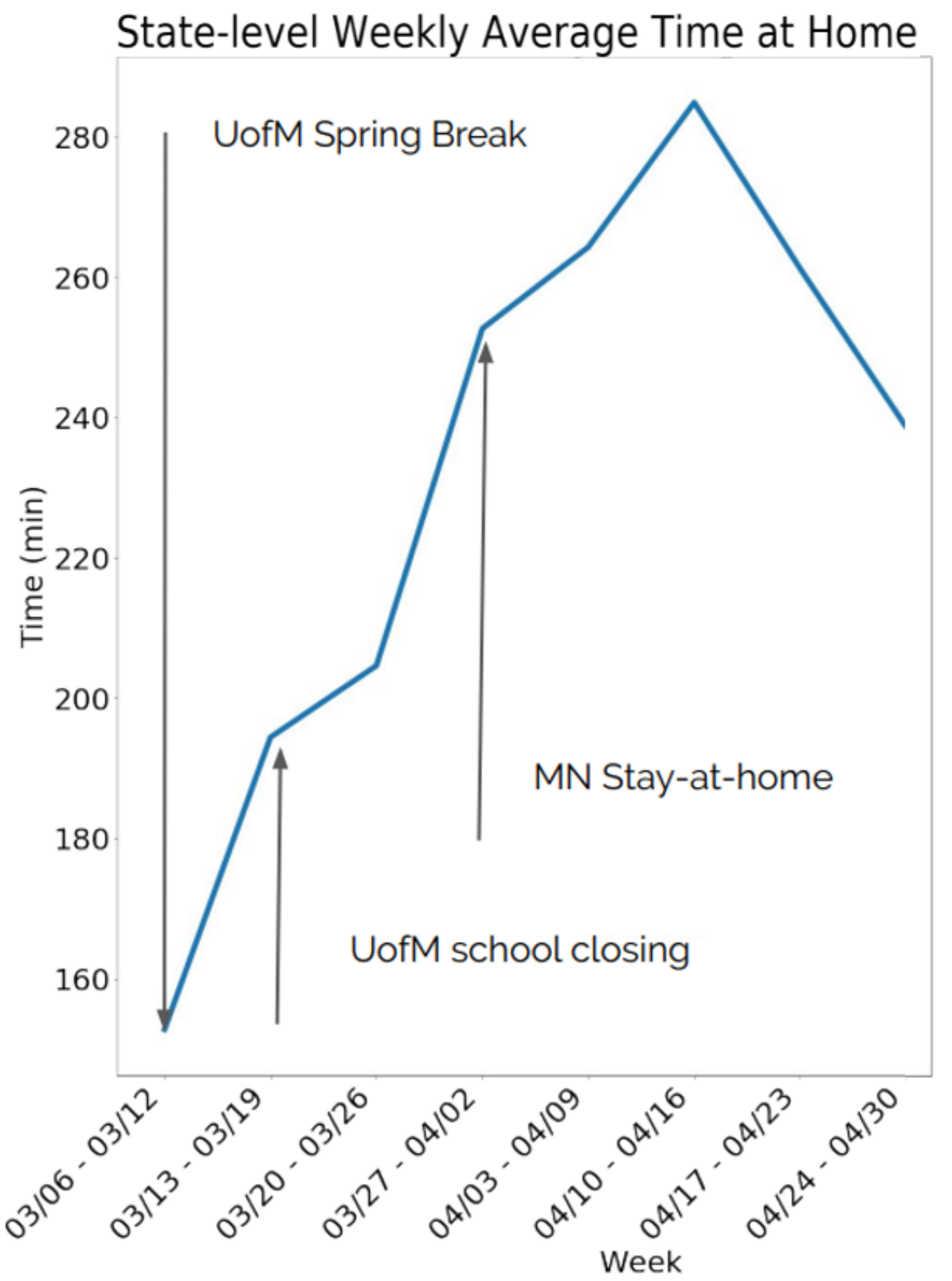}
        \caption{Weekly Average Time at Home\\}
        \label{fig:avg_time_home}
     \end{subfigure}
        \caption{Mobility assessment in MN State between 03/02/2020 - 04/30/2020}
        \label{MobilityAssesment}
\end{figure*}

Besides information value and report quality, we also provide a brief justification on how the proposed community-engaged decision support system addresses problems beyond the limited questions of specific community stakeholders presented in Figure 1. We further analyzed \textbf{compliance behavior towards stay at home (SAH) order} generated by the proposed system. The state government issued Stay at Home orders from $27^{th}$ March 2020 - $18^{th}$ May 2020, which included the closing of bars and restaurants \cite{SAH1}. Figure \ref{fig:avg_dist_from_home} shows a significant drop in average distance traveled from home until April 2020, after which we see an increase in mobility activity. This accords with the trend we see in Figure \ref{fig:avg_time_home}, which shows an increase in Average Time at Home until 10th April, followed by a decrease in later weeks.
In addition, the stay-at-home order was followed by an extended stay-at-home order \cite{SAH1}. This order called for the gradual reopening of businesses after June 10, when full-service restaurants could operate at a limited indoor capacity. The result was an increase in foot traffic, as shown in Figure 7. To monitor adherence to the extended stay-at-home on order hourly or daily basis, information on the number of long (i.e., greater than 20 mins) duration visits to certain businesses (e.g., full-service restaurants) is necessary (as shown in Figure 8-10). A public API (Application Programming Interface) which reports these values at regular intervals can serve this need as described in Section 4.3.

\section{Other Related Work}\label{related_work}
A web-based mapping platform \cite{dong2020interactive} by Johns Hopkins University was the earliest work in response to tracking the COVID-19 outbreak in real-time. A pioneer web-based that was developed to provide reported daily information, including the total number of confirmed cases, recovered, and death on COVID-19 across the entire globe. Following the same line of work, other surveillance applications and web-based platforms \cite {gao2020mapping, samet2020using, desjardins2020rapid} were developed to visualize and provide real-time quantification data, such as mobility changes and emerging hotspots. One platform incorporates GIS and daily human mobility statistical trends (e.g., median travel distance) derived from anonymized and aggregated mobile location big data at the county level in the United States \cite{gao2020mapping}. Another web-based application, CoronaViz \cite{samet2020using}, visualizes COVID-19 spread across the globe using animation, which allows users to change the spatial region and time span interval. A surveillance application is described in the literature \cite{desjardins2020rapid} that uses space-time scan statistics \cite{kulldorff2005space} to detect emerging hotspots on a daily basis. It then calculates the relative risk score at the county-level and finally visualizes the results on a US map. While all of these web-based map visualizations are useful for understanding the COVID-19 spread at a low spatial scale (i.e., county), the main component of engaging end-users (e.g., policymakers) and closing the loop by delivering customized reports is missing. Customized reports can improve decision-making and allocation of resources.

Another line of research that has attracted a lot of attention during the past year is contact tracing applications. Many studies \cite {park2020contact, aleta2020modelling, kretzschmar2020impact} have demonstrated the promise of contact tracing for safe re-opening of the economy and business following state and federal policy interventions in response to the COVID-19 crisis. However, the effectiveness of this technology to help with the ongoing pandemic and possible future outbreaks is depend on its accessibility and usage by a large population at a fine-grained geographical scale \cite {mokbel2020contact}. Several factors influence accessibility, including socioeconomic conditions, age, and smartphone penetration. Another issue that has been extensively discussed in both the media and academia is privacy concerns, which deter people from participating in contact tracing applications. Interested readers can refer to a recent survey of automatic contact tracing approaches using Bluetooth low energy for more details \cite{reichert2021survey}. Other application of contact tracing may also include interpreting missing segment via space-time modeling of trajectory signal gap \cite{sharma2020analyzing} or time-series weather forecasting  \cite{sharma2018webgiobe}.

Similar to \cite{chang2021mobility}, the authors in \cite{nouvellet2021reduction} develop a framework to infer the relationship between mobility and the population-level transmission by defining a parametric relationship between transmission and mobility using the effective reproduction number as a parameter and evaluated by fitting two models to the country-specific time-series of COVID-19 deaths. In \cite{huang2020understanding}\cite{kang2020multiscale} discusses the change in human mobility based on transportation-related behaviors (e.g., modes of transportation, etc.) and their spatial interactions patterns at different scales (e.g., census tracts, county, and state level) via dynamically inferring Origin-Destination flows using US Census data. In \cite{coston2021leveraging}, the authors link smart-phone based data (Safegraph) to high-fidelity ground truth administrative data. This enables auditing mobility data for bias in the absence of demographic information and ground truth labels. This illuminates demographic disparities and how such disparities distort policy decisions.

Beyond contact tracing applications, two popular disease dynamic transition models, namely SIR (susceptible, infectious, and recovered) and SEIR (susceptible, exposed, infected, and recovered,) have recently been adapted to benefit from spatiotemporal data to simulate the spread of COVID-19. Bobashev et al. \cite{bobashev2020geospatial} proposed a unified framework to combine data-driven predictive modeling (i.e., reinforcement learning) with overall trajectories of disease dynamics from a mechanistic model (i.e., SEIR model) for forecasting COVID-19 spread. A variant of the SIR model, namely the time-dependent SIR model \cite{chen2020time} models transmission rate as well as recovery rate at a given time \emph{t}. This work also considers both how infected but asymptomatic individuals may contribute to the spread of COVID-19 disease. However, the primary limitations of variant compartmental models (i.e., SIR, SEIR) are sensitive to rapid transmission and high fluctuations in true positive tests (i.e., infected individual data), which are not available at the individual level. In addition, these models largely depend on certain assumptions to estimate the total population. They do not explicitly account for the changes in mobility induced by commuting and short trips, which necessitates additional post-processing to calibrate the model.

\section{Conclusion and Future Work}\label{C_FW}
In this work, we proposed a community-engaged COVID-19 decision support system which addresses queries (e.g., long-duration visits) posed by end-users. The architecture provides custom reports related to user specific queries related to actual decision making questions which are not addressed by state of the art dashboards. To address such queries, we designed an Entity-Relationship diagram on weekly pattern data from the SafeGraph dataset. The new schema can enhance our understanding of the Safegraph data and analyze what additional data need to be integrated to improve support of ad-hoc queries. Finally, we validate the proposed decision support system on a real-world mobility data with a case study of Minnesota and briefly summarize end-user feedback along with a discussion of data quality concerns.

\textbf{Future Work:} In future work, we are interested in investigating part-time and full-time employment trends and modeling the mobility impacts with respect to those attributes. In addition, we will investigate data quality by further characterizing the relationship between disease spread and mobile-phone data. We will also address the effect of 3rd normal schemas on data storage cost, processing time, choice of indexing and query processing strategies. We will also integrate a real-time data stream in the proposed system. In addition, we will compare Safegraph data trends with other data sources (e.g., Foursquare) to asses data quality issues. Finally, we will explore a new spatial data mining technique that can automate routine tasks and further consolidate it with an interest measure that can distinguish hangout places, emerging hotspots, and static hotspots.

\section*{Acknowledgement}
This material is based upon work supported by the National Science Foundation under Grants No. 2040459 and 1737633.  We would also like to thank our collaborators from the data provider (SafeGraph) to MIT media lab (Dr. Esteban Moro Egido and Dr. Alex Pentland), Metropolitan council (Dr. Ashley Asmus), State of Minnesota (MnDOT, MMB, and MN DHS), and the University of Minnesota (Center for Transportation Studies (Laurie McGinnis), School of Public Health (Dr. Eva Enns, Dr. Shalini Kulasingam, and Dr. Kelly Searle), U-Spatial (Len Kne, Adam Null), Metro Transit (Dr. Eric Lind) and Dept. of Industrial and Systems Eng. (Dr. Ankur Mani) for their insightful comments and constructive guidelines. We also want to thank Kim Koffolt and the spatial computing research group for their helpful comments and refinements.

\bibliographystyle{ACM-Reference-Format}
\bibliography{manuscript.bib}

\end{document}